\documentclass[onecolumn]{aa} % for a paper on 1 column  
\usepackage{natbib}
\usepackage{graphicx}
\usepackage{color}
\usepackage{amsmath}
\usepackage{float}
\usepackage{longtable}
\usepackage[toc,page]{appendix}
\usepackage{txfonts}
\begin{document}

 \title{The A-shell star $\phi$ Leo revisited: its photospheric and
     circumstellar spectra\thanks{Table C.1, and the spectra used
     in this paper are only available in electronic form at the CDS
     via anonymous ftp to cdsarc.u-strasbg.fr (130.79.128.5)
     or via http://cdsweb.u-strasbg.fr/cgi-bin/qcat?J/A+A/}}

\author{C. Eiroa
     \inst{1}\fnmsep\thanks{Part of this work was made while at the Departamento de F\'\i sica Te\'orica of the Universidad Aut\'onoma de Madrid, and as visiting astronomer at Calar Alto Observatory.}
     \and B. Montesinos\inst{2}
     \and I. Rebollido \inst{3}
     \and Th. Henning \inst{4}     
     \and R. Launhardt \inst{4}
     \and J. Maldonado\inst{5} 
      \and G. Meeus \inst{6}      
      \and A. Mora \inst{7}
      \and P. Rivi\`ere-Marichalar \inst{8}
      \and E. Villaver \inst{2}
}

      %1
      \institute{Private Researcher %\email{no-gmail}
      \and
      %2
      Centro de Astrobiolog\'{\i}a (CAB, CSIC-INTA), ESAC Campus, s/n, 28692
      Villanueva de la Ca\~nada, Madrid, Spain\\
      \email{bmm@cab.inta.csic.es}
      \and
      %3
      Space Telescope Science Institute, 3700 San Martin Drive, Baltimore,
      MD 21218, USA
      \and
      %4
      Max-Planck-Institut f\"ur Astronomie (MPIA), K\"onigstuhl 17, D-69117
      Heidelberg, Germany
      \and
      %5
      INAF, Osservatorio Astronomico di Palermo, Piazza del Parlamento 1, 90134 Palermo, Italy
      \and
      %6
      Departamento de F\'{\i}sica Te\'orica, Universidad Aut\'onoma de Madrid, 28049 Madrid, Spain
      \and
      %7
      Aurora Technology B.V. for ESA, ESA-ESAC, 28692 Villanueva de la Ca\~nada, Madrid, Spain
      \and
      %8
      Observatorio Astron\'omico Nacional (OAN,IGN), Calle Alfonso XII 3, 28014 Madrid, Spain
}

   \date{}
 
  \abstract
  % context heading (optional)
  % {} leave it empty if necessary
  {Variable red- and blue-shifted absorption features observed in the 
    Ca {\sc ii} K line towards the A-type shell star $\phi$ Leo have
    been suggested by us in a previous work to be likely due to solid,
    comet-like bodies in the circumstellar (CS) environment.}
  % aims heading (mandatory)
  {Our aim is to expand our observational study of this object to other
    characteristic spectral lines of A-type photospheres as well as to lines
    arising in their CS shells.}
  % methods heading (mandatory)
  {We have obtained more than 500 high-resolution optical spectra
    collected at different telescopes during 37 nights in several
    observing runs from December 2015 to January 2019. Time series
    consecutive spectra were taken, covering intervals of up to $\sim$9
    hours in some nights. We have analysed some photospheric lines,
    in particular Ca {\sc i} 4226 \AA ~and Mg {\sc ii} 4481 \AA, as
    well as the circumstellar shell lines Ca {\sc ii} H\&K, Ca {\sc ii}
    IR triplet, Fe {\sc ii} 
     4924, 5018 and 5169 \AA,
    Ti {\sc ii} 3685, 3759 and 3761 \AA, and the Balmer lines H$\alpha$
    and H$\beta$.}
  % results heading (mandatory)
  {Our observational study reveals that $\phi$ Leo is a variable
    $\delta$ Scuti star whose spectra show remarkable dumps and bumps superimposed 
    on the photospheric line profiles, which vary their 
    strength and sharpness, propagate from blue- to more red-shifted radial
    velocities and persisting during a few hours. Similarly to other
    $\delta$ Scuti stars, these features are likely produced by
    non-radial pulsations. At the same time, all shell lines present
    an emission at $\sim$3 km/s centered at the core of the CS
    features, and two variable absorption minima at both sides of the
    emission; those absorption minima occur at practically the same
    velocity for each line, i.e., no apparent dynamical evolution is
    observed. The variations observed in the Ca {\sc ii} H\&K, Fe 
    {\sc ii} and Ti {\sc ii} lines occur at any time scale from minutes
    to days and observing run, but without any clear
    correlation or recognizable temporal pattern among the different
    lines. In the case of H$\alpha$ the CS contribution is also variable
    in just one of the observing runs.}
  % conclusions heading (optional), leave it empty if necessary
  {Summarizing, we suggest that $\phi$ Leo is a rapidly rotating
    $\delta$ Scuti star surrounded by a variable, (nearly) edge-on CS
    disk possibly re-supplied by the $\delta$ Scuti pulsations. 
    The behaviour of the CS shell lines is reminiscent of the one 
    observed in rapidly rotating Be shell stars with an edge-on CS disk, 
    and clearly points out that the variations observed in the CS features 
    of $\phi$ Leo are highly unlikely to be produced by exocomets. In
    addition, the observational results presented in this work,
    together with some recent results concerning the shell star HR 10,
    suggest the need of a critical revision of the Ca {\sc ii} K
    features which have been attributed to exocomets in other shell
    stars. }

   \keywords{ stars: individual: $\phi$ Leo -- Stars: early-type --
     stars: variables: $\delta$ Scuti -- stars: Be -- circumstellar
     matter -- comets: general }

   \maketitle

\section{Introduction}

Transient red-  and blue-shifted, absorption features  mainly detected
in the  Ca {\sc  ii} K  line towards $\sim$30  A-type stars  have been
interpreted  as the  signatures  of the  evaporation  of large  solid,
comet-like  bodies  transiting  or   grazing  onto  de  central  stars
\citep[e.g.][]{ferlet87,kiefer14b,welsh18}.  The  interpretation about
the presence  of such  bodies is  reinforced by  the detection  in the
surroundings of  a few of  those stars  of sporadic events  and stable
absorptions in  some UV metallic lines,  as well as emission  lines at
far-IR  and (sub)-mm  wavelengths indicating  the presence  of
secondary cold  gas  in  some  debris  disk  stars.   Additionally, 
photometric observations have revealed the presence of exocomets around
later type stars \citep{boyajian16,kiefer17,rappaport18};
further, photometric and spectroscopic evidences of such bodies
are   found   around white dwarf stars  \citep[see e.g.][and references
  therein]{manser19,strom20}.

Several of the exocomet host A-type stars are known to be shell stars. 
Those objects -whose CS envelope is most likely a CS disk due to their high rotational 
velocities ($v \sin i$) - have variable, narrow absorption features in many lines of e.g. 
Ti {\sc ii}, Fe {\sc ii}, and also the Ca {\sc ii} H\&K and infrared (IR) triplet lines, 
with the shell lines even appearing or disappearing in some stars \citep[e.g.][]{jaschek88,abt97}. 
However,  the  similar variability behaviour of the \mbox{Ca {\sc  ii} K}  
line between the shell stars and the bona-fide cometary-like events detected 
towards other A-type stars like $\beta$ Pic or  HD 172555  \citep[see e.g.][]{kiefer14a,kiefer14b} can  
lead to some  misidentification of  the shell  variability (or  star 
binarity) with events from exocomets.  A  paradigmatic case of this 
confusion is provided  by  the  shell star  HR  10  where  the  variability 
of  the  narrow absorption components superimposed  on the photospheric 
Ca  {\sc ii} K line, as well as on Ti {\sc  ii} absorption lines at 
3759/61 \AA ~were interpreted  as  comet-like  events caused  by  
infalling,  comet-like material onto the star
\citep{lagrangehenri90,welsh98,redfield07a,abt08}.             However,
\cite{montesinos19}, based on  the analysis of a large  set of optical
spectra  spanning  several   decades,  together  with  interferometric
PIONIER/VLTI data,  showed that the spectroscopic  variations observed
in HR  10 are actually caused  by its binary nature  with CS envelopes
around both stellar components.

In this work we revisit the shell star $\phi$ Leo. In a previous paper
\citep{eiroa16} we analysed  the Ca {\sc ii}  H\&K non-photospheric CS
absorptions  and  attributed  the observed,  striking  variability  to
comet-like events as the most  plausible explanation --although we note
that   a  question   mark  was   included   in  the   title  of   that
paper. However, the analysis of  other lines and numerous time series
spectra taken in 2017 and  2019, presented here, brings new perspectives
about the star's variability, and we deem now our previous analysis as
inappropriate.

This paper is organized as  follows: Sect. 2 summarises the properties
of the  star,  Sect.  3  describes the observations, instruments and
observing  periods,  Sect. 4 describes  the results  concerning some
photospheric and non-photospheric CS  lines, with specific examples of
spectra obtained in some periods, in Sect.  5 we present a discussion
of these results, and Sect. 6 is a brief summary of our conclusions.

\section{$\phi$ Leo: properties}
\label{hr4368}

$\phi$  Leo   (HD  98058,  HR   4368)  is  an  A5V-A7 IVn   shell  star
\citep[e.g.][]{jaschek91} located at a distance  of 56.5 pc.  The best
stellar  parameters found  by \cite{eiroa16}  fitting Kurucz  photospheric 
models to high  resolution spectra  are $T_{\rm  eff}$ =  7500 K,
$\log g$  = 3.75, $v\sin i$  = 230 km/s, in  reasonable agreement with
other estimates \citep[e.g.][]{lagrangehenri90,royer07, zorec12, adamczak14,david15,soubiran16}.
Its    bolometric    luminosity    is    $\sim$45--50    L$_\odot$
\citep{zorec12,adamczak14,balona20}; the  estimated mass  is $\sim$2.3
M$_\odot$ \citep{zorec12,adamczak14,derosa14}, while age estimates are
in  the  range  $\sim$500-900 Myr  \citep{derosa14,david15},  and  the
stellar radius  has been  estimated to be  $\sim$3.2 $R_\odot$  with a
significant  oblateness  \citep{vanbelle12,arcos18}.   $\phi$  Leo  is
located in  the $\delta$  Scuti instability strip  of the  HR diagram,
and has been classified  by  \cite{balona20}  as  a $\delta$  Scuti  variable  with
several  peaks in  its {\em TESS}  periodogram,  the strongest  one with  an
amplitude of 5.25 ppt and frequency of 6.4739 d$^{-1}$.

Shell Ca  {\sc ii}  H\&K and  Ti {\sc ii}  3759, 3761  \AA ~absorption
lines   have   been  detected   in   the   spectrum  of   $\phi$   Leo
\citep[e.g.][]{abt73,jaschek88}.
The shell  lines likely arise in  a gaseous, close to  edge-on CS disk
given  the  high  rotational velocity of the  star  \citep{abt08};  thus,  the
triangular shape of the  Ca {\sc ii} K profile is  probably due to the
combination    of    the    photospheric    and    disk    absorptions
\citep{lagrangehenri90}.  Long term  variations of the Ti  {\sc ii} CS
lines were reported by \cite{abt08}, while we found remarkable short term
variations of the Ca {\sc ii} H\&K lines \citep{eiroa16}.  Finally, we
note that  the star does  not possess a  detectable warm or  cold dusty
debris disk \citep{rieke05,cataldi19}.

\begin{table*}[t]
\caption{Log of observations}
\small
\begin{tabular}{llllll}
  \hline
  \hline
  \noalign{\smallskip}
Observing Run & Instrument & Range(nm) & Resolution& Dates$^\dagger$ & Spectra per night \\
\noalign{\smallskip}
\hline
\noalign{\smallskip}
2015 December& HERMES     & $\sim$370-900&$\sim$85000&20, 22, 23     & 1, 4, 1           \\
2016 January & FIES       & $\sim$370-830& $\sim$67000&26             & 2                 \\
2016 January & HERMES     &        &      &27, 28, 30     & 1, 3, 3           \\
2016 March   & HERMES     &        &      &3, 4, 5, 6     & 4, 4 ,2 ,3        \\
2016 March   & FEROS      & $\sim$350-930& $\sim$48000&25, 26, 27, 28 & 4, 4, 3, 3        \\
2016 May     & HERMES     &        &       &11             & 20                \\
2016 May     & CARMENES   & $\sim$520-960/960-1710&$\sim$94600/80400&19             & 1                 \\
2017 March   & HERMES     &                 &           &6, 7, 8, 9,    & 14, 20, 22, 26,   \\
             &            &                 &           &10, 11, 12, 13  & 26, 13, 24, 24     \\
2017 March/April & HERMES &                 &           &28, 29, 30, 31,& 24, 24, 26, 33,   \\
                 &        &                 &           &1, 2, 3        & 21, 22, 23        \\
2017 March/April& FEROS   &                 &           &31, 1, 2, 3,      & 8, 28, 26, 11,     \\ 
                 &        &                 &           &4, 5, 6, 7, 8      & 16, 16, 13, 16, 15 \\
2019 January & HARPS-N    & $\sim$392-683         &$\sim$115000      &29             & 13                \\ 
\noalign{\smallskip}
\hline
\noalign{\smallskip}
\multicolumn{6}{l}{$\dagger$ Dates refer to the beginning of the night,
  this convention is used throughout the paper.} 
\end{tabular}
\label{table:log}
\end{table*}

\section{Observations}

A total of 555 high-resolution spectra of $\phi$ Leo were
collected during 10 observing campaigns from December 2015 to April
2017: 389 spectra were obtained with HERMES \citep{hermes} attached 
to the Mercator Telescope (La Palma, Spain); 163 spectra with FEROS
\citep{feros} at the MPG/ESO 2.2-m telescope (La Silla, Chile); two
spectra were obtained with the FIES spectrograph \citep{fies} at NOT
(La Palma, Spain), and one more with CARMENES \citep{carmenes}, the
high-resolution optical/near-IR spectrograph attached at the 3.5-m
telescope at Calar Alto observatory (Almer\'\i a, Spain). Additionally,
13 spectra were taken in January 2019 with HARPS-N \citep{cosentino12}
at TNG (La Palma).

Observing dates, instruments - with their corresponding spectral 
range and resolution - and number of spectra per night are given
in Table \ref{table:log}; a detailed log with the specific dates and
time (UT) is given in Table C.1 of Appendix
\ref{Appendix_Observing_Log}\footnote{For the sake of compactness,
the dates and times in the table are codified as
yyyymmddThhmm where yyyy, mm and dd are year, month and day, respectively, 
and hh mm are the UT hour and minute of the beginning of the observation. 
The table is ordered from left to right and top to bottom.}. 
During the 2017 observing runs $\phi$ Leo was extensively observed when
its airmass was $\lesssim 2$; in particular, from March 31 to April 3 the
star was continuously followed from La Palma and La Silla along $\sim$9.5
hours.

The spectra were reduced using the available pipelines of the
corresponding instruments; they include the usual steps for \'echelle
spectra, as bias subtraction, flat-field correction, cosmic ray
removal, and order extraction.  Wavelength calibration is carried out
by means of Th-Ar lamp spectra taken at the beginning and end of each
night.  Barycentric corrections have been applied to the HERMES, FIES,
and CARMENES spectra since the pipelines do not include such
correction.  Finally, telluric lines have been removed by means of
MOLECFIT, a tool that generates a model atmosphere accounting for the
most common absorbing molecules \citep{smette15}.  Typical S/N ratios
of the spectra achieved at a continuum around 4000 \AA~, i.e., close
to the Ca {\sc ii} H\&K lines, are in the range $\sim$70--150, mostly
depending on weather conditions. S/N ratios of the median spectra of
each night are in the range $\sim$200-500, again depending on weather
conditions and the number of spectra per night, and the one of the
median spectrum of all spectra is $\sim$1000.

\begin{figure*}[!ht]
\centering
\scalebox{0.30}{\includegraphics[angle=-90]{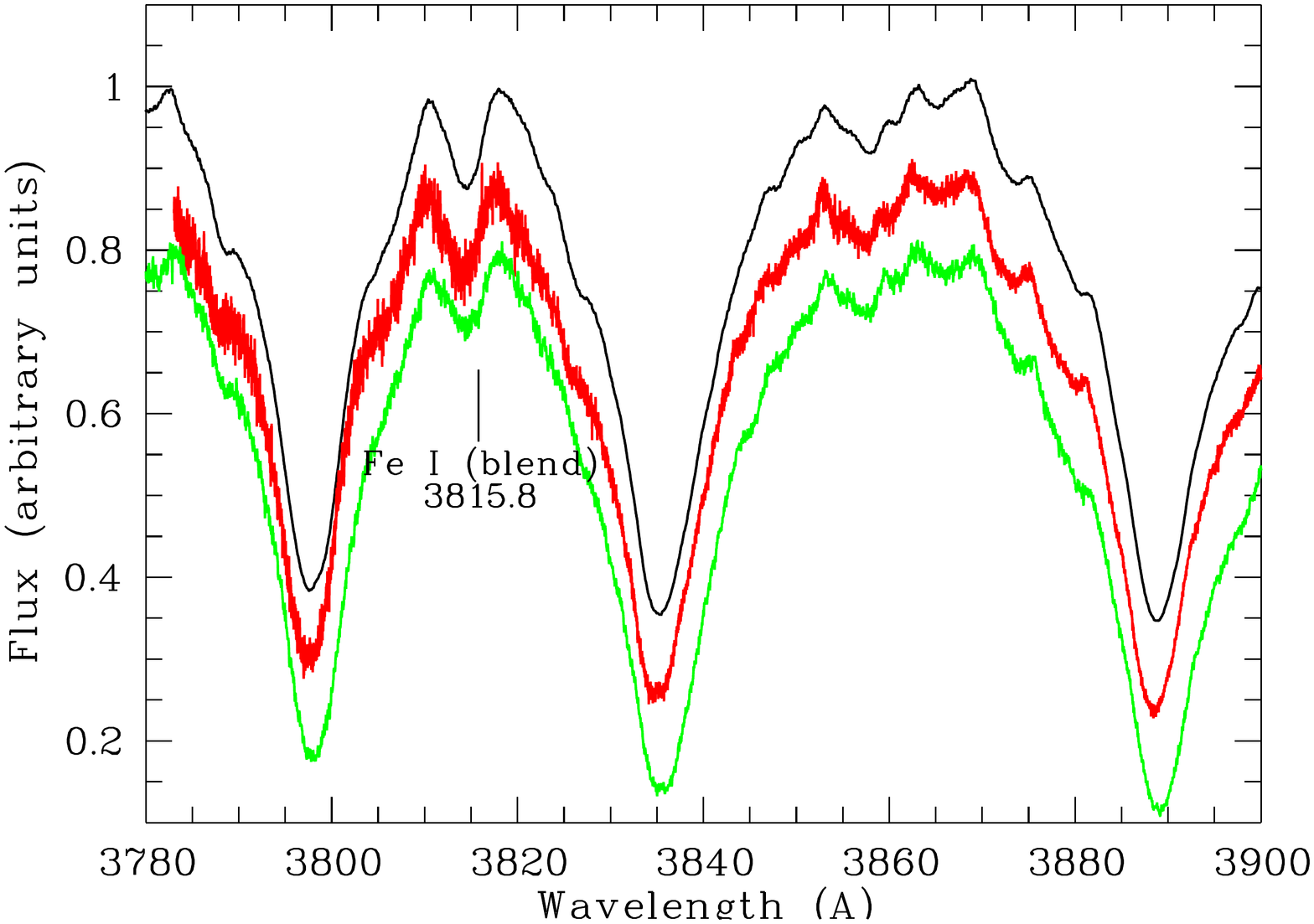}}
\scalebox{0.30}{\includegraphics[angle=-90]{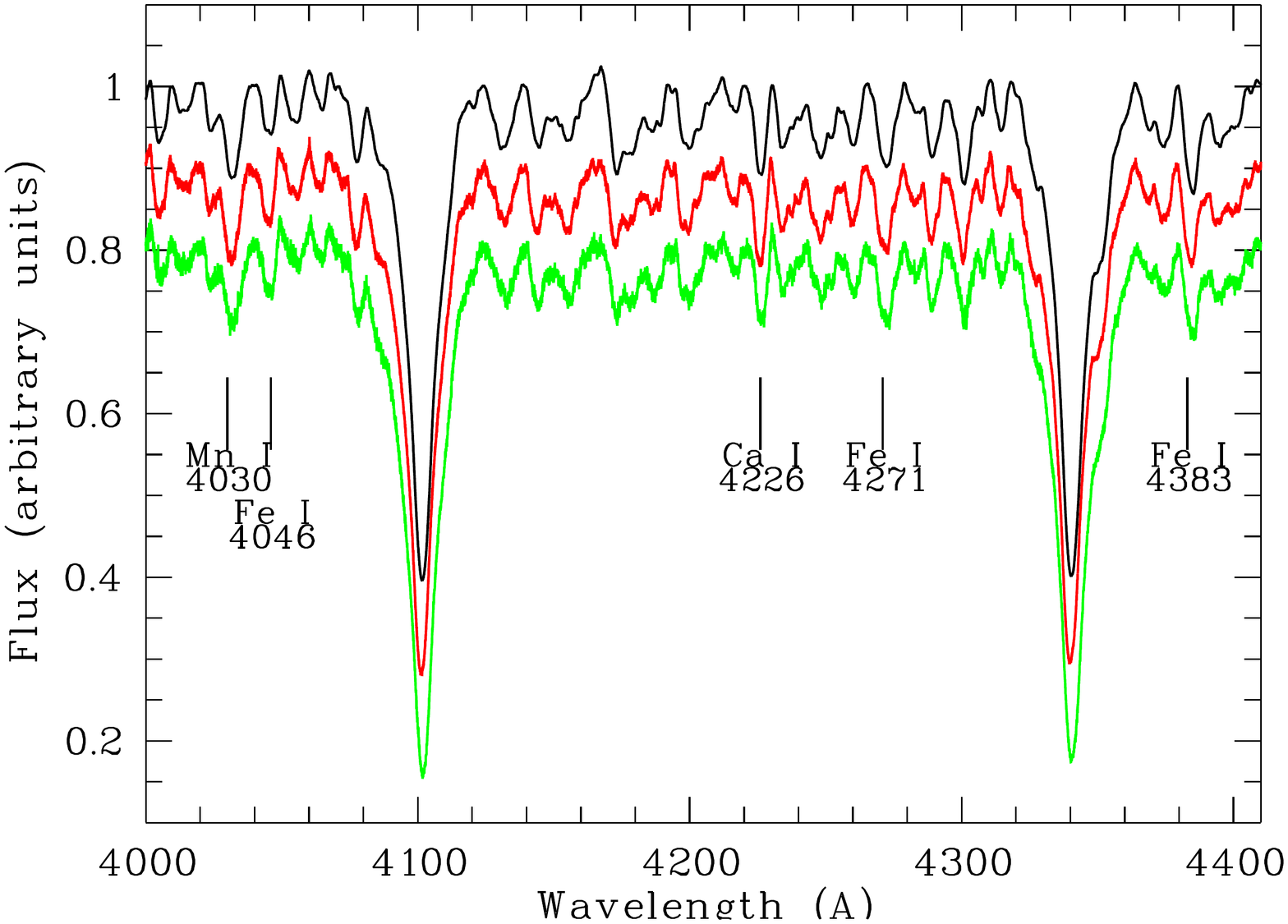}}
\scalebox{0.30}{\includegraphics[angle=-90]{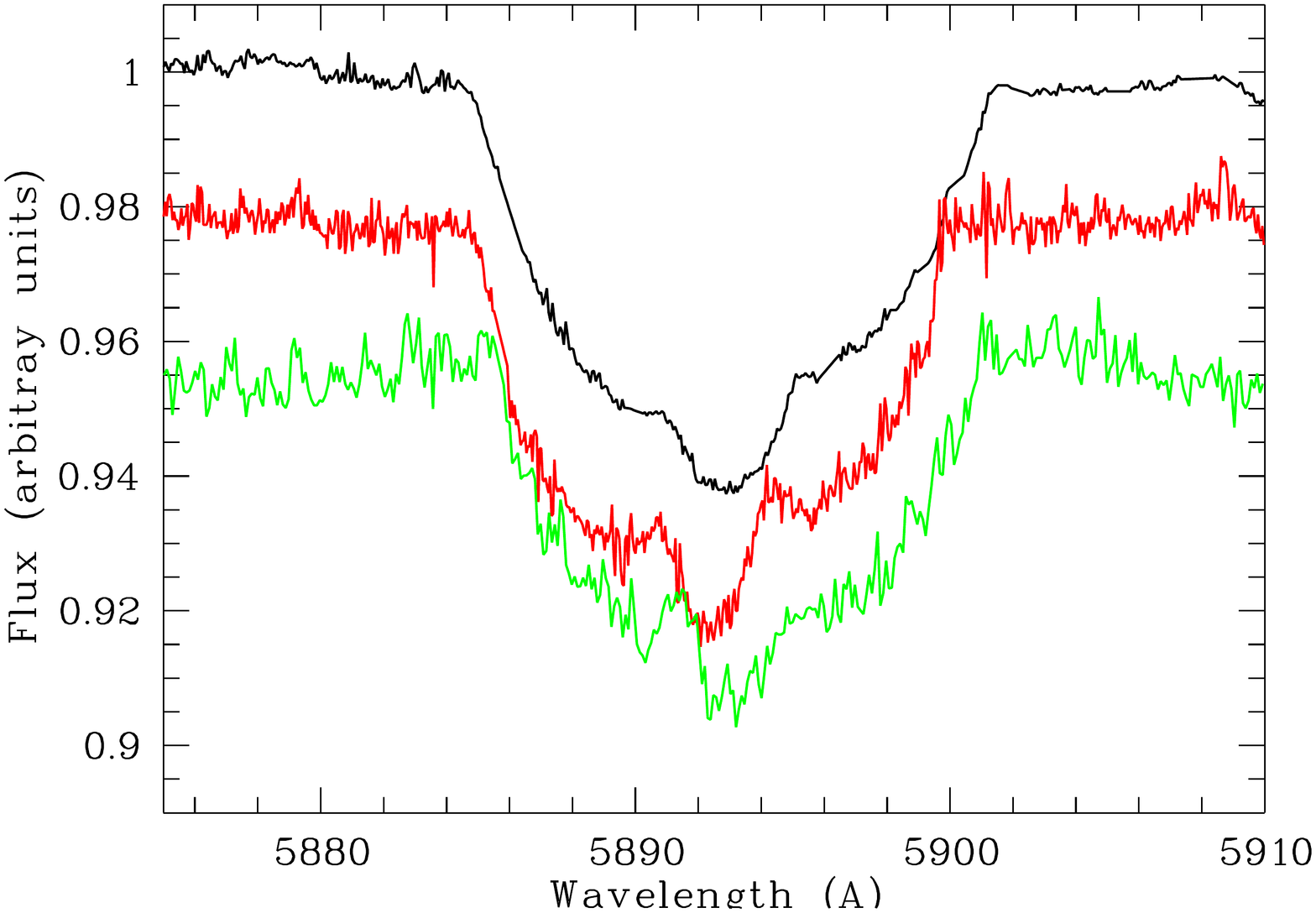}}
\scalebox{0.30}{\includegraphics[angle=-90]{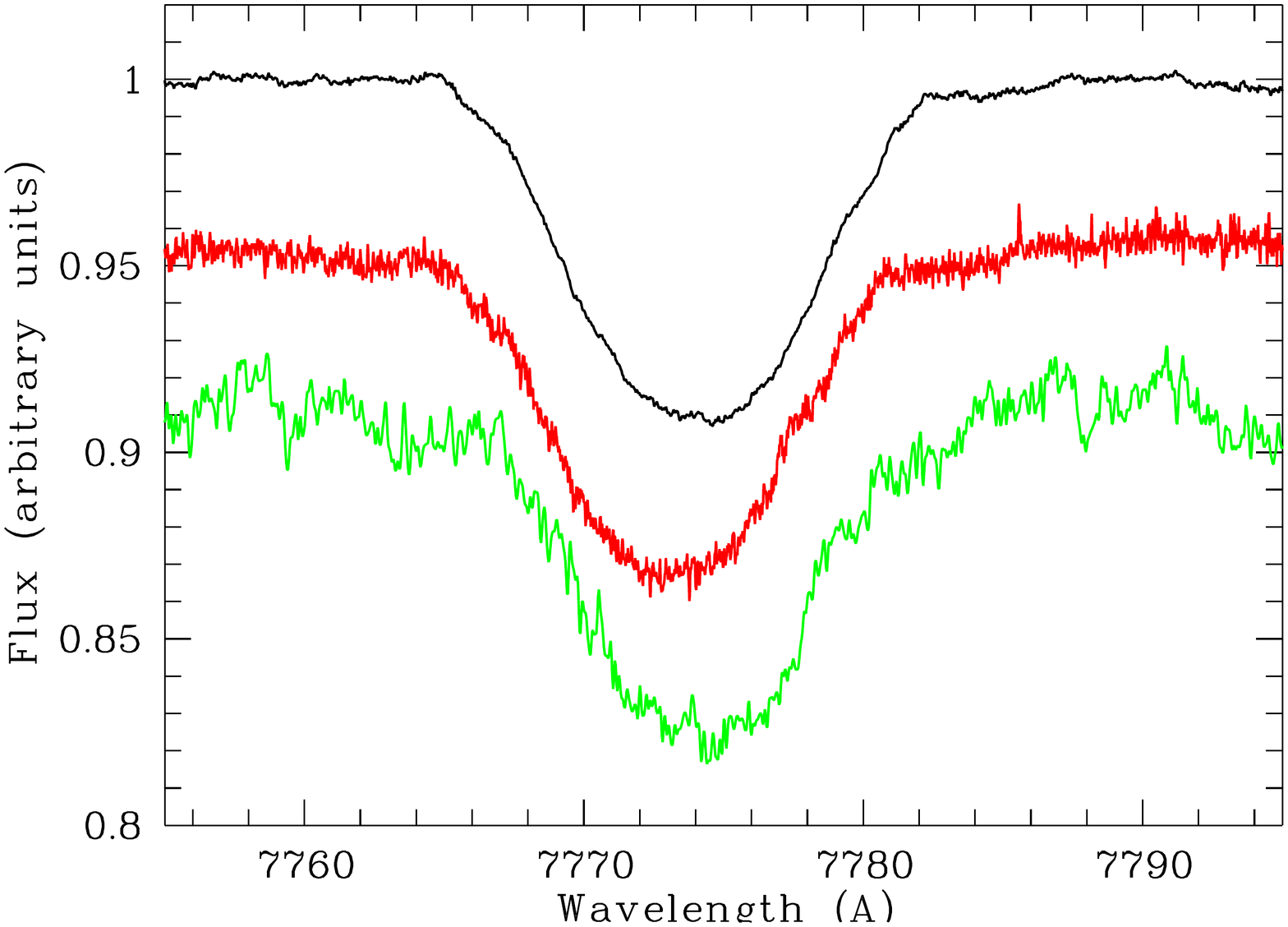}}
\caption{Photospheric lines of $\phi$  Leo (black).  Top: Balmer lines
  together with  some lines  due to neutral  atoms. The  absorption at
  $\sim$ 3815  \AA ~is  most likely a  blend of Fe  {\sc i}  lines (the
  referenced wavelength denotes  the strongest Fe {\sc i}  line of the
  blend, i.e. the  one with the highest oscillator  strength). Bottom:
  Na {\sc ii}  D 5890/96 \AA  ~and  O {\sc i}  triplet at 7775
  \AA.  For  comparison the  spectra  of  $\alpha$  Aql (red)  and 
  $\kappa$ Phe  (green) are  also plotted. Strength  and width  of the
  Balmer lines  and the rest of  the spectral features are  similar in
  all  three  stars indicating  that  they  are ``pure''  photospheric
  absorptions.}
\label{photospheric} 
\end{figure*}

\begin{figure*}[!ht]
\centering
\scalebox{0.30}{\includegraphics[angle=-90]{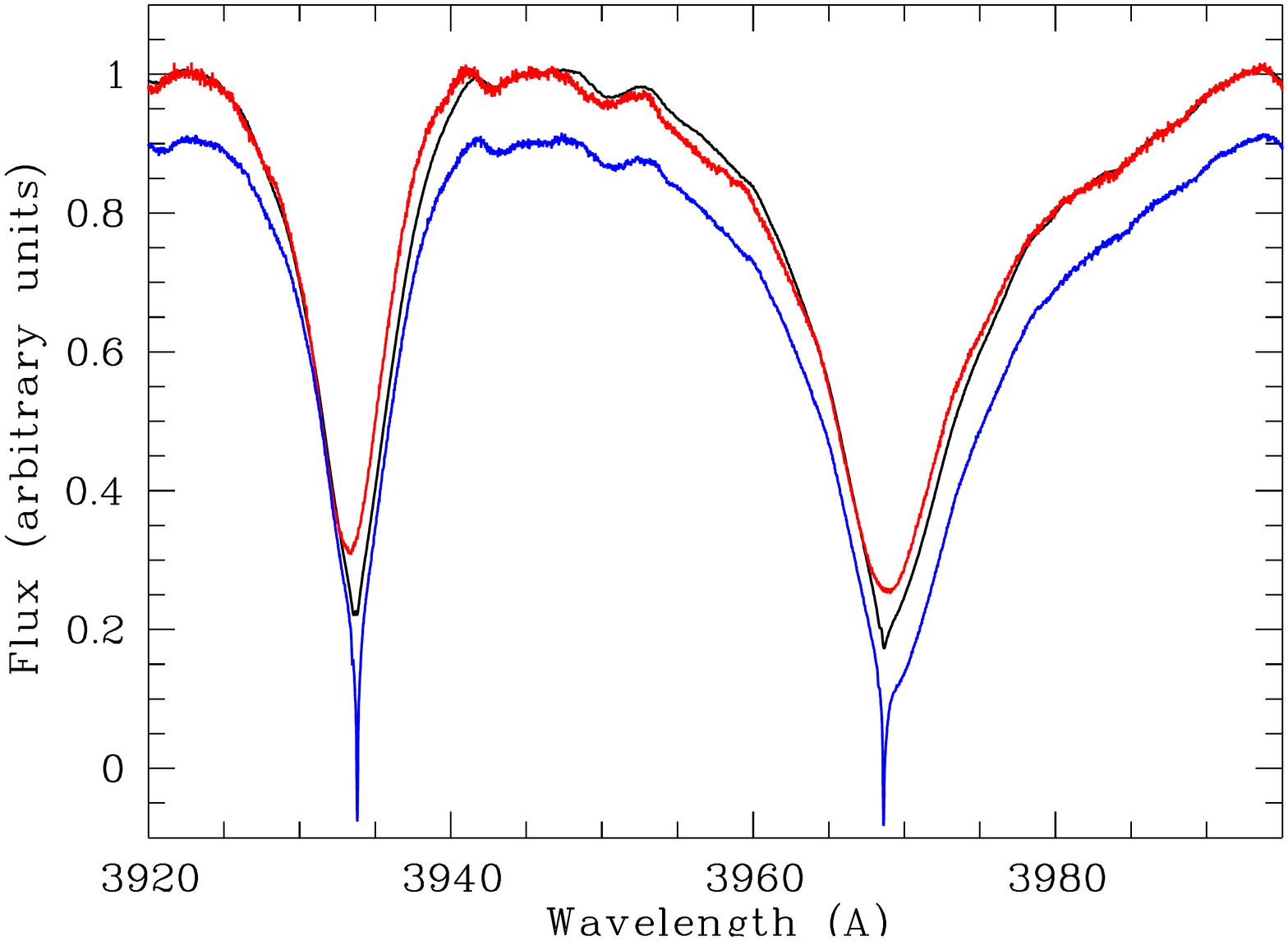}}  
\scalebox{0.30}{\includegraphics[angle=-90]{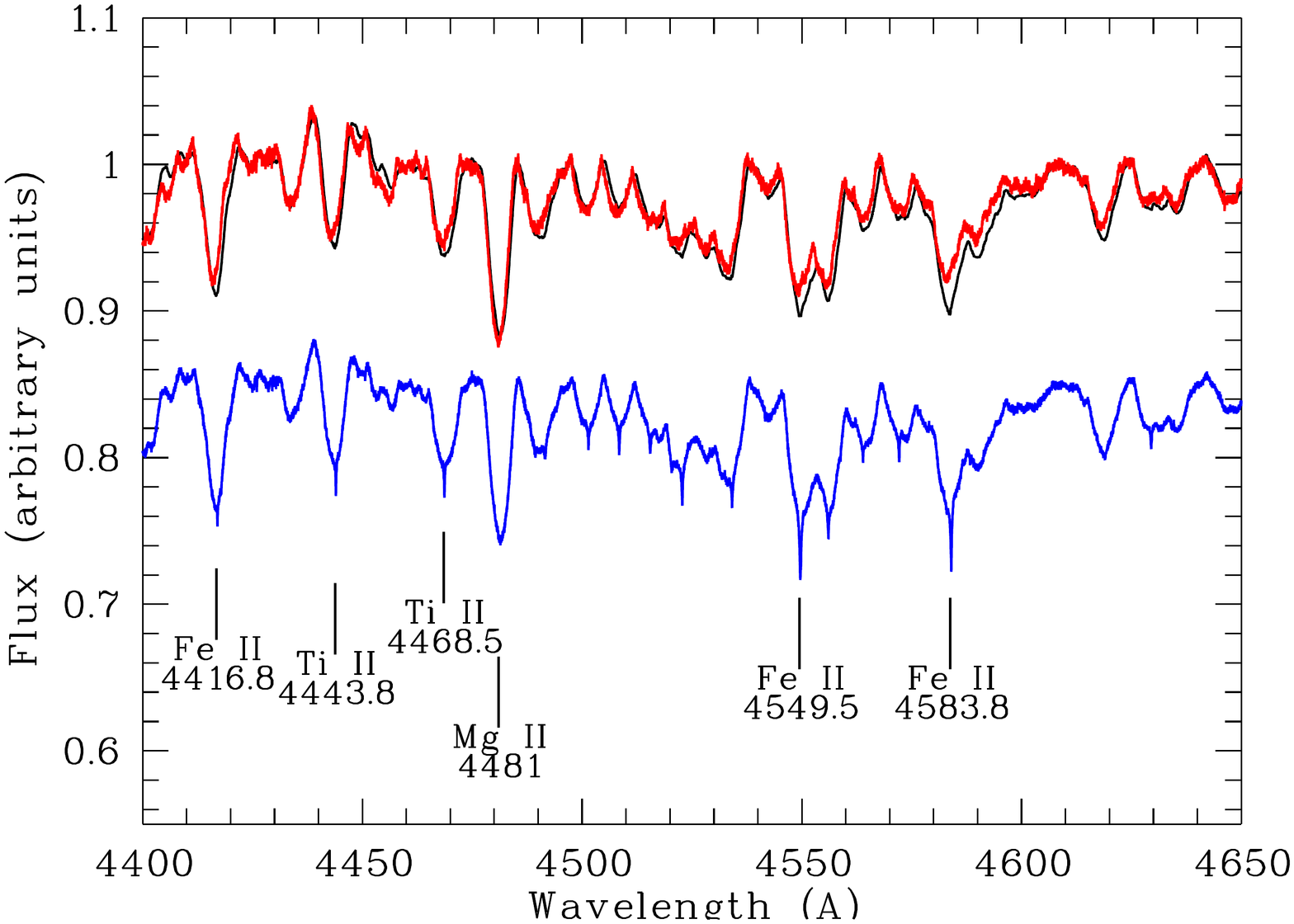}}
\scalebox{0.30}{\includegraphics[angle=-90]{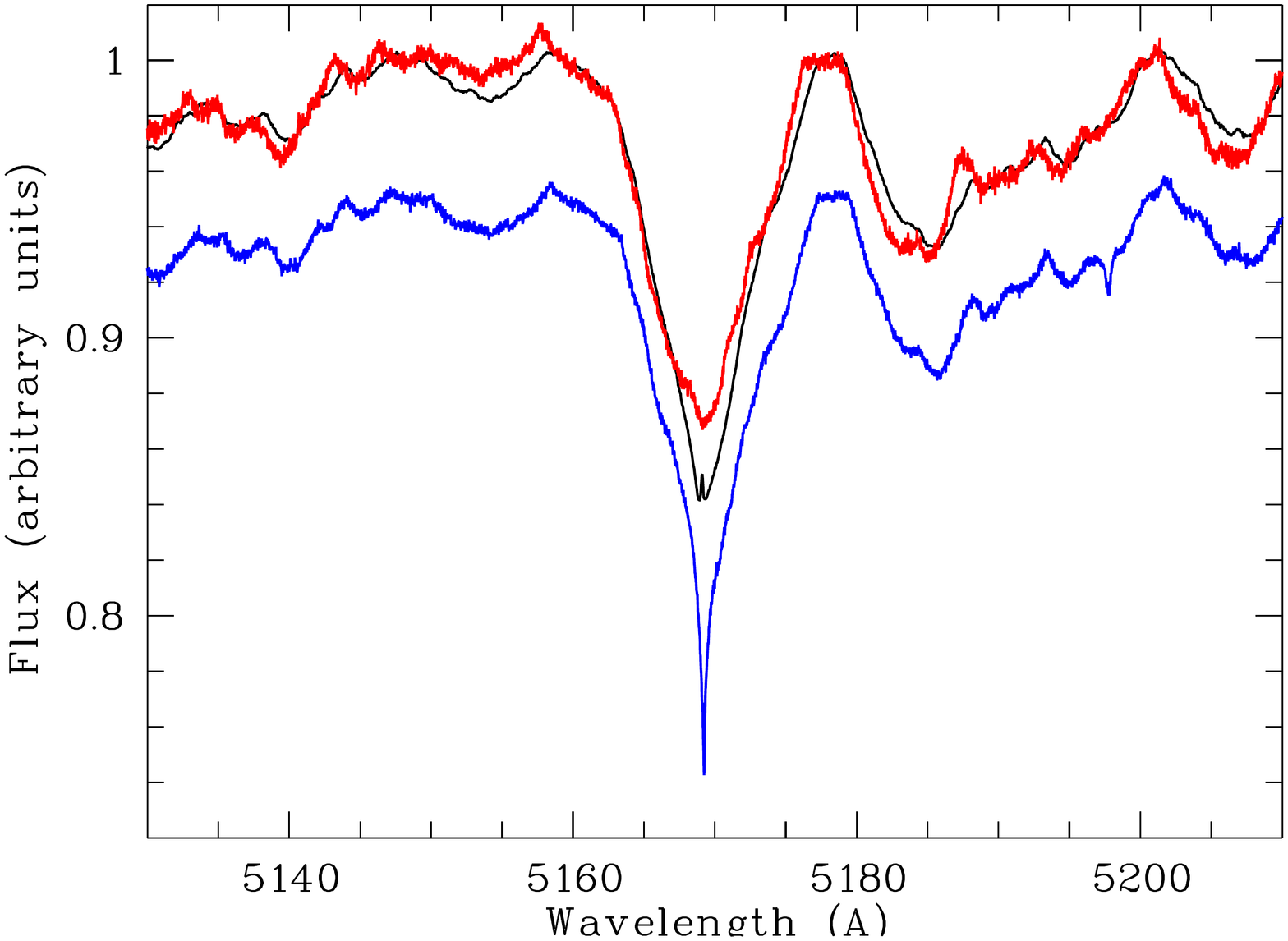}}
\scalebox{0.30}{\includegraphics[angle=-90]{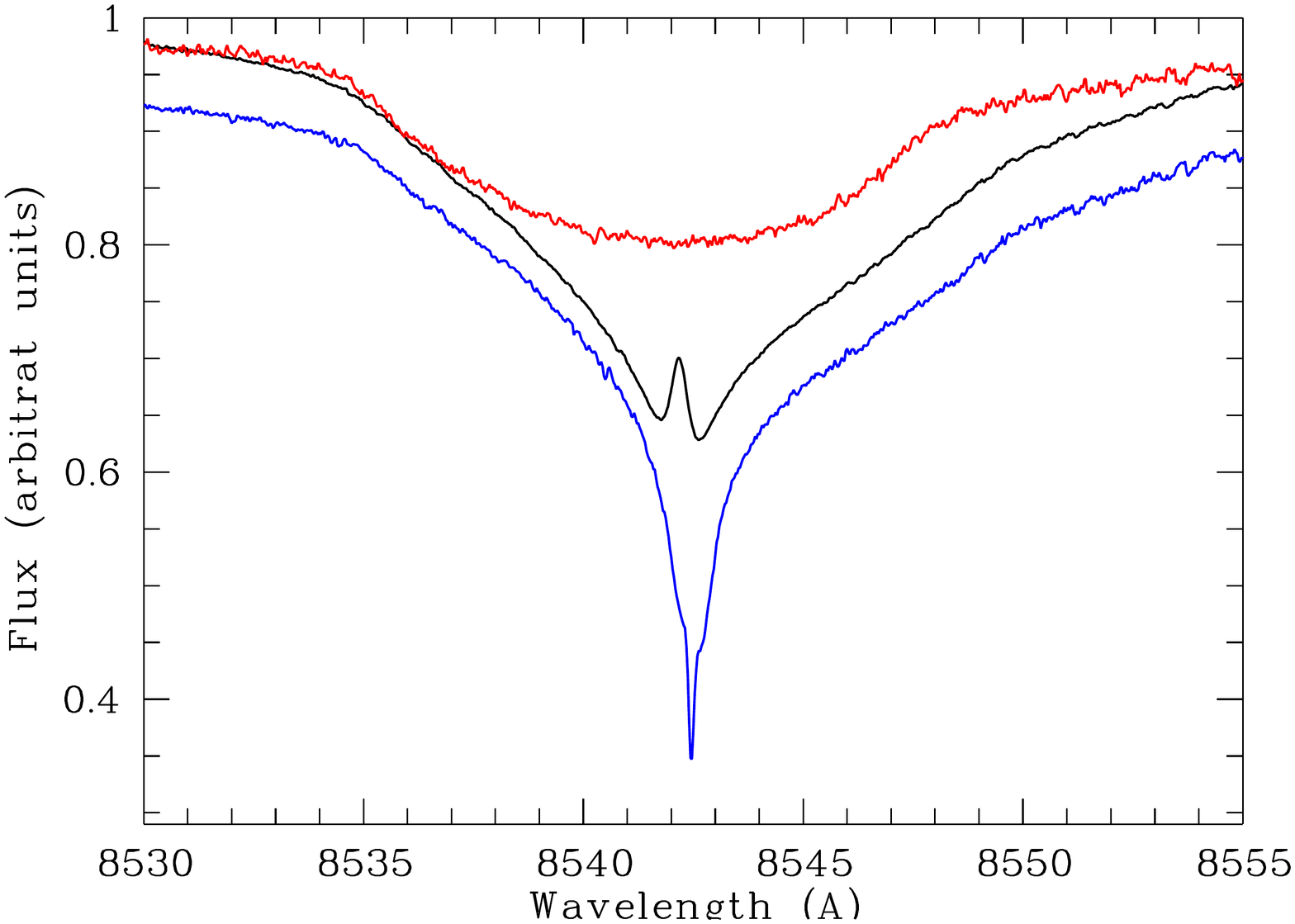}}
\caption{Top Left: Ca  {\sc ii} H\&K lines and  the Balmer H$\epsilon$
  of  $\phi$ Leo  (black), together  with $\alpha$  Aql (red)  and the
  shell  star   HR  7731  (blue).  The   conspicuous  non-photospheric
  contribution in the Ca {\sc ii} lines of $\phi$ Leo is distinguishable
  although not as prominent as in  HR 7731.  Top Right: the same three
  stars. In this  case the photospheric line Mg {\sc  ii} 4481 line is
  similar in all stars, while the  shell Fe {\sc ii} and Ti {\sc
    ii} lines  are striking  in HR  7731 but not  that much  in $\phi$
  Leo. Bottom  left: the  5169 \AA  ~line of the  Fe {\sc  ii} triplet
  42. Bottom right: the Ca {\sc ii} triplet line at 8542 \AA.  The spectrum of  $\alpha$  Aql is plotted superimposed on the $\phi$ Leo spectrum in all panels.}
\label{nonphotospheric} 
\end{figure*}

\begin{figure}[!ht] 
\centering
\scalebox{0.26}{\includegraphics[angle=-90]{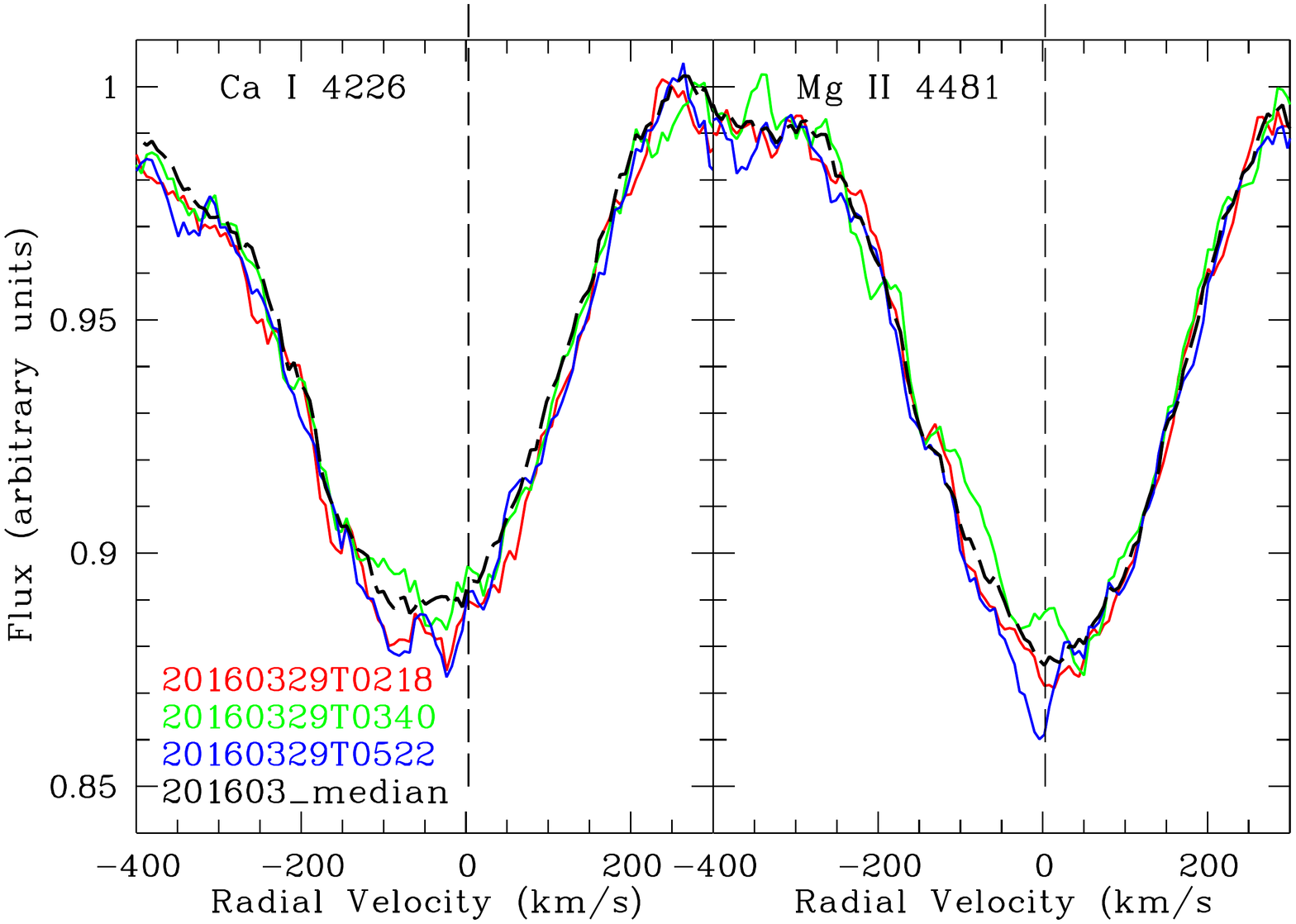}}
\caption{Ca {\sc i} 4226 \AA ~and Mg {\sc ii} 4481 \AA ~lines observed
  with FEROS during  the night of 2016-03-28/29.  The  spectra have been
  rebinned to a pixel spectral  resolution of 6.39 km/s.  Date and UT
  of each spectrum  are indicated. Black dashed line: median  of the 4
  day period shown for comparison. Note the different behaviour of the extra absorption 
  features of each line when compared to their median spectra.  The vertical dashed line corresponds to a  radial velocity  of 3 km/s (see text in subsection \ref{sec:variabilityofcsabsorptions} and Fig. \ref{cs:velocities_median}).}
\label{201603_cai_mgii}  
\end{figure}

\section{Results: The spectrum of $\phi$ Leo}

The spectrum of $\phi$ Leo consists of a broadened A-type photospheric
spectrum together with some  CS contribution.  Fig. \ref{photospheric}
shows the median spectrum of $\phi$ Leo in some wavelength ranges with
characteristic photospheric Balmer and neutral metallic lines (Fe {\sc
  i}, Mn {\sc  i}, Ca {\sc i}, Na  {\sc i}, O {\sc i})  of late-type A
stars \citep[e.g.][]{gray09}.  This median spectrum has been estimated
by using  all FEROS and  HERMES spectra -the latter have been
rebinned to the FEROS spectral resolution (0.03 \AA/pixel).  The
spectra of  $\alpha$ Aql and  $\kappa$ Phe, two stars  with values of
T$_{\rm eff}$ and $ v\sin i$ comparable to those of $\phi$ Leo 
\citep{rebollido20}, are also shown with the aim  of comparison.  The 
strength and width of the photospheric lines are similar in all three stars .

In addition, $\phi$  Leo exhibits a number of absorption  lines with a
clear  non-photospheric   contribution.   Fig.   \ref{nonphotospheric}
shows  examples of  such lines;  in particular  the Ca  {\sc ii}  H\&K
lines, some Fe {\sc ii} and Ti  {\sc ii} lines around the 4481 \AA ~Mg
{\sc ii} photospheric one, the \mbox{Fe {\sc  ii} 5169 \AA}, and the 8542 \AA
~line  of the  Ca {\sc  ii} IR triplet.  In  this figure the spectrum  of
$\alpha$  Aql  is plotted  directly  superimposed  on the  $\phi$  Leo
spectrum.  Fig.  \ref{nonphotospheric} additionally displays the spectrum 
of the shell star HR  7731  which has also  similar photospheric  
parameters \citep{rebollido20} .  We note that the strength of some shell 
lines in $\phi$ Leo might  be a bit stronger than in  $\alpha$ Aql, e.g. 
the Fe  {\sc  ii} 4549.5, 4583.8  \AA ~lines,  but  not
comparable with the very strong shell absorptions observed in HR 7731.

$\phi$  Leo   exhibits  conspicuous  photospheric   and  circumstellar
variability  as evidenced  by the  behaviour of  distinctive lines  of
A-type  shell   stars.  The  photospheric  line   variability  becomes
discernible as dumps  and bumps in the line profiles  with similar but
not  identical  characteristics  in   different  lines; a careful 
inspection of several regions close to the photospheric lines, shows that 
such bumps and dumps are not seen in the continuum. The  temporal presence 
of  such features  is irregular and,  when they  appear, they usually 
move from blue  to red in short time scales, lasting seemingly for up to a few hours. On the other hand, the CS lines show a common  
emission at their core, and absorption changes  within very short time 
scales at both  the blue and red sides of the core,  yet without any  
apparent correlation among the different lines.  In the next two 
subsections we give  further  details of  the photospheric  and  
CS  line  variability separatedly.

\subsection{Photospheric lines}

\begin{table}[!ht]
  \caption{Equivalent Widths (m\AA) of Ca {\sc i}  4226 \AA ~ and Mg {\sc ii} 4481 \AA ~during the night 2016-03-28/29.}
\label{tab:EWs_mg_ca}

\centering
\begin{tabular}{ccc}
\hline \hline
\noalign{\smallskip}
UT (hh:mm)  &  Ca {\sc i}    & Mg {\sc ii}  \\
\noalign{\smallskip}
\hline
\noalign{\smallskip}
02:18     &   534 &    528   \\
03:40     &   511   &   501\\
05:22     &  550    &   554\\
median    & 530   &    528 \\
\noalign{\smallskip}
\hline
\noalign{\smallskip}
\multicolumn{3}{l}{Estimated uncertainty $\sim5$\%}
\end{tabular}
\end{table} 

\begin{figure*}[!ht] 
  \centering
\scalebox{0.30}{\includegraphics[angle=-90]{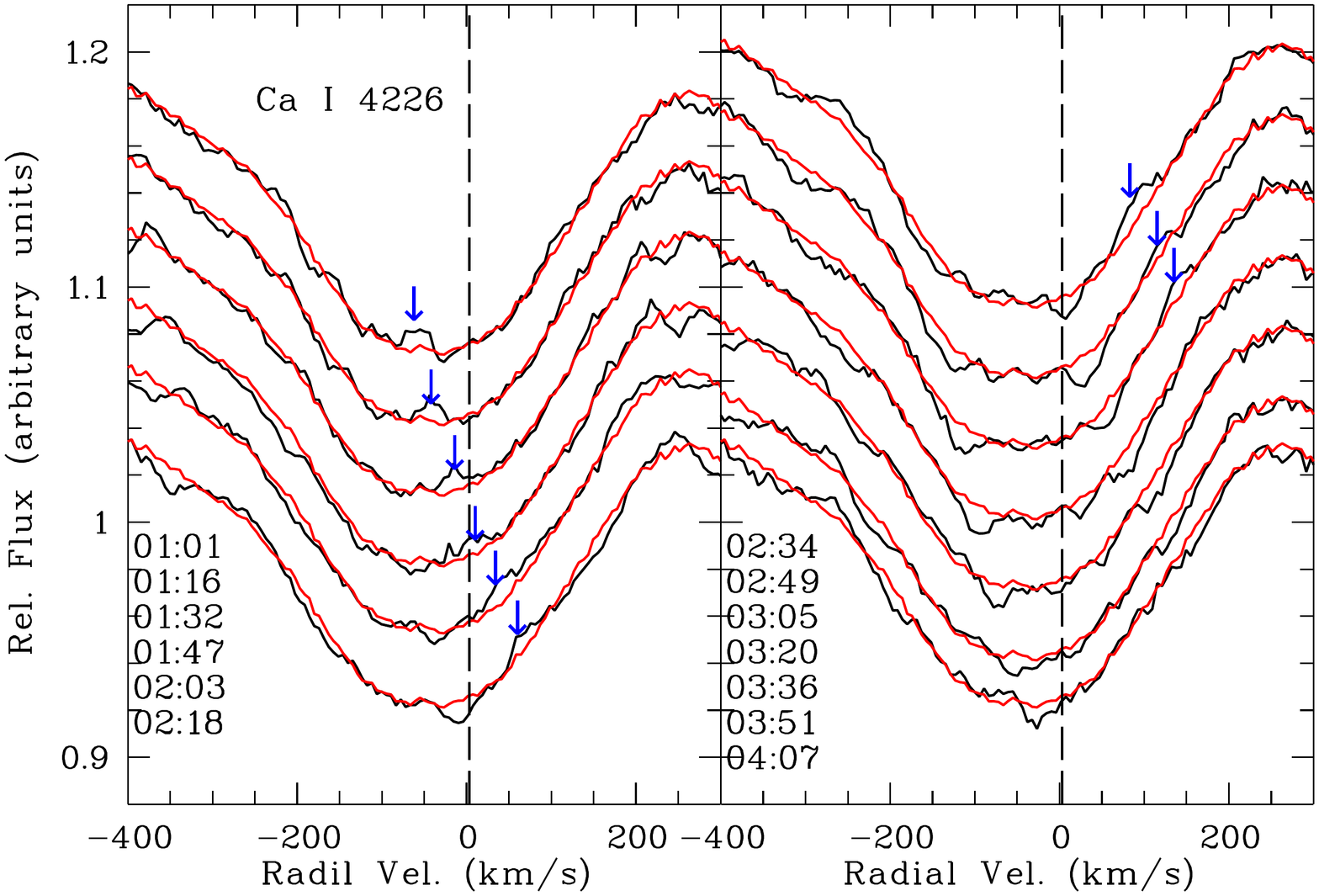}}
\scalebox{0.30}{\includegraphics[angle=-90]{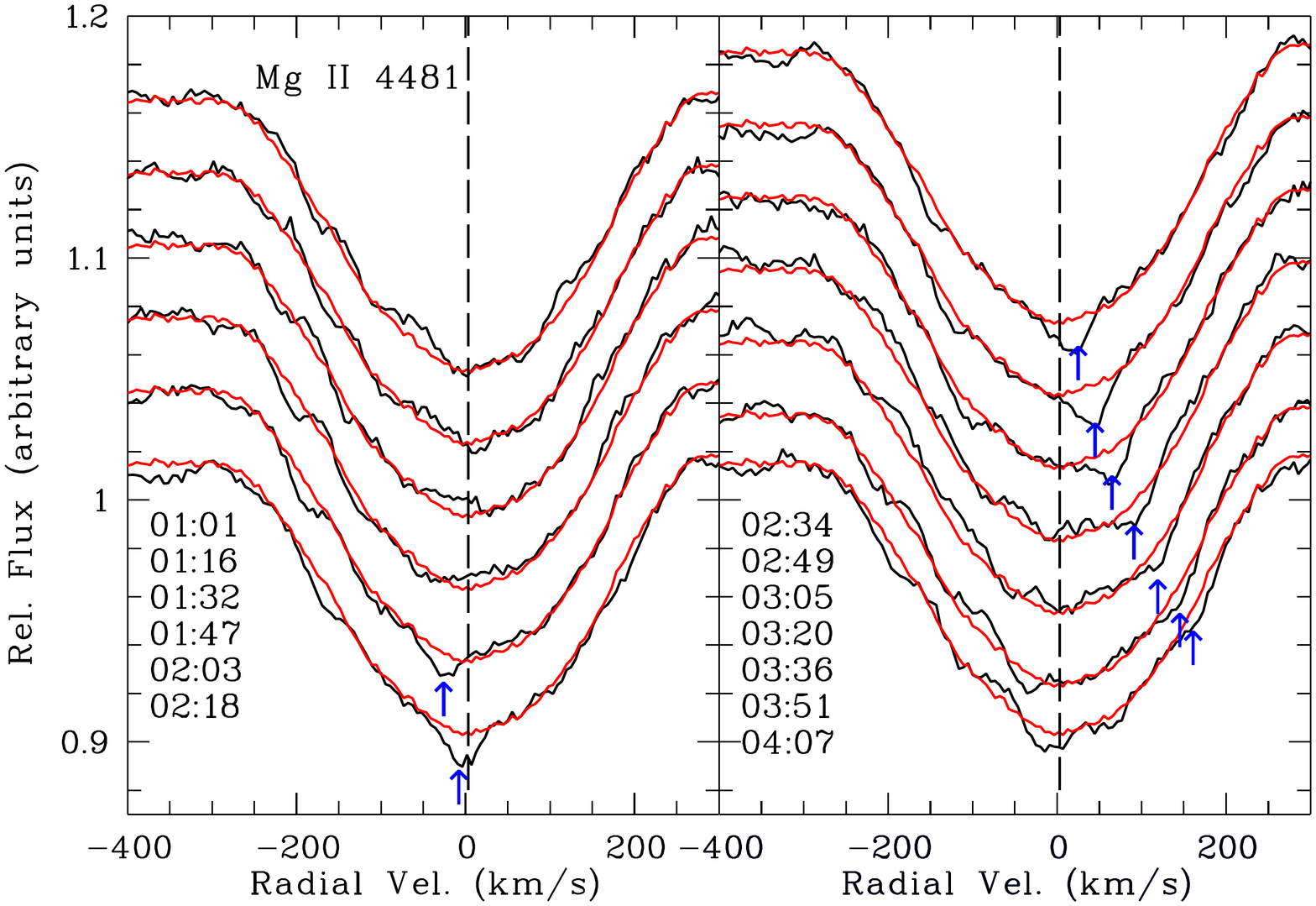}}
\scalebox{0.30}{\includegraphics[angle=-90]{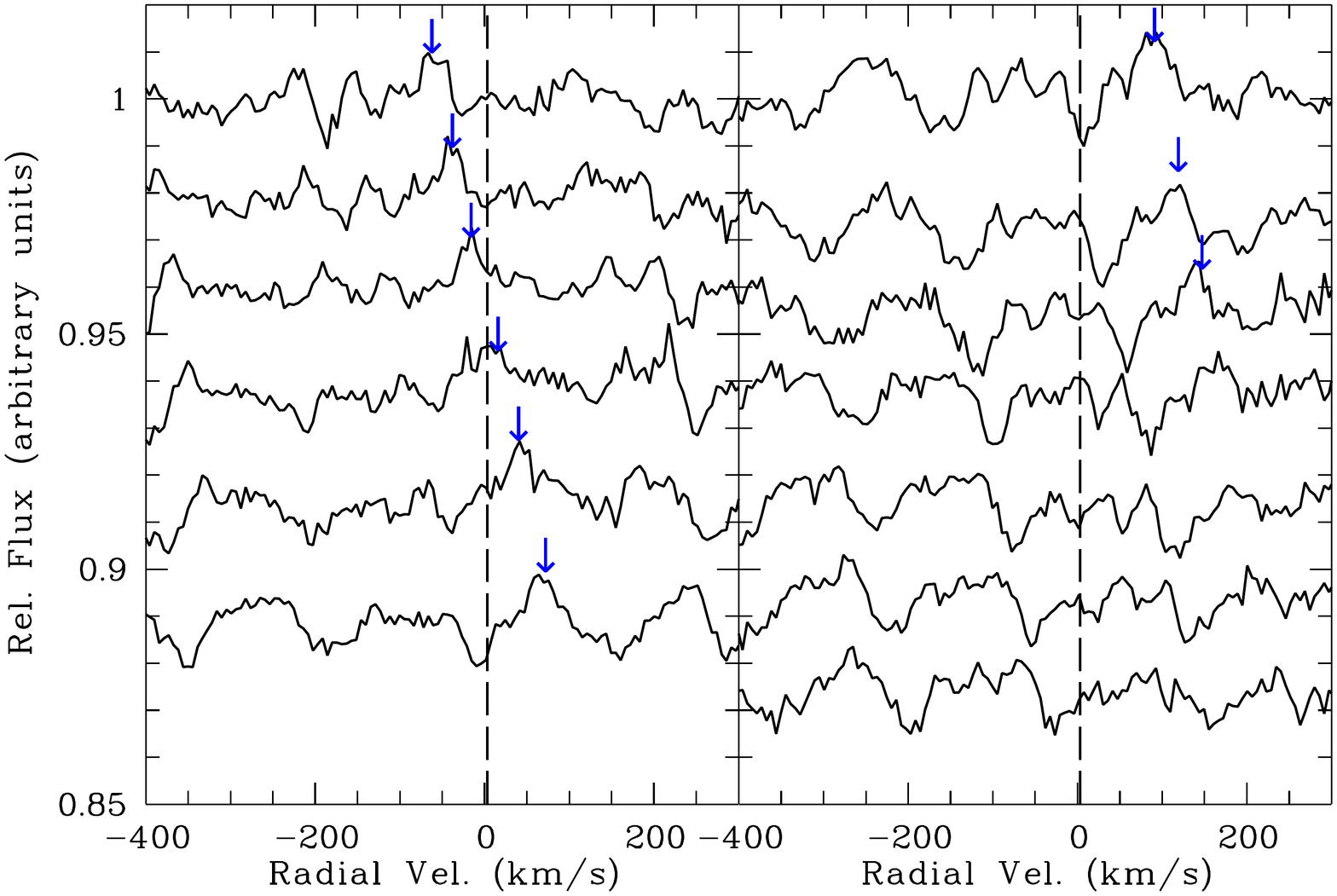}}
\scalebox{0.30}{\includegraphics[angle=-90]{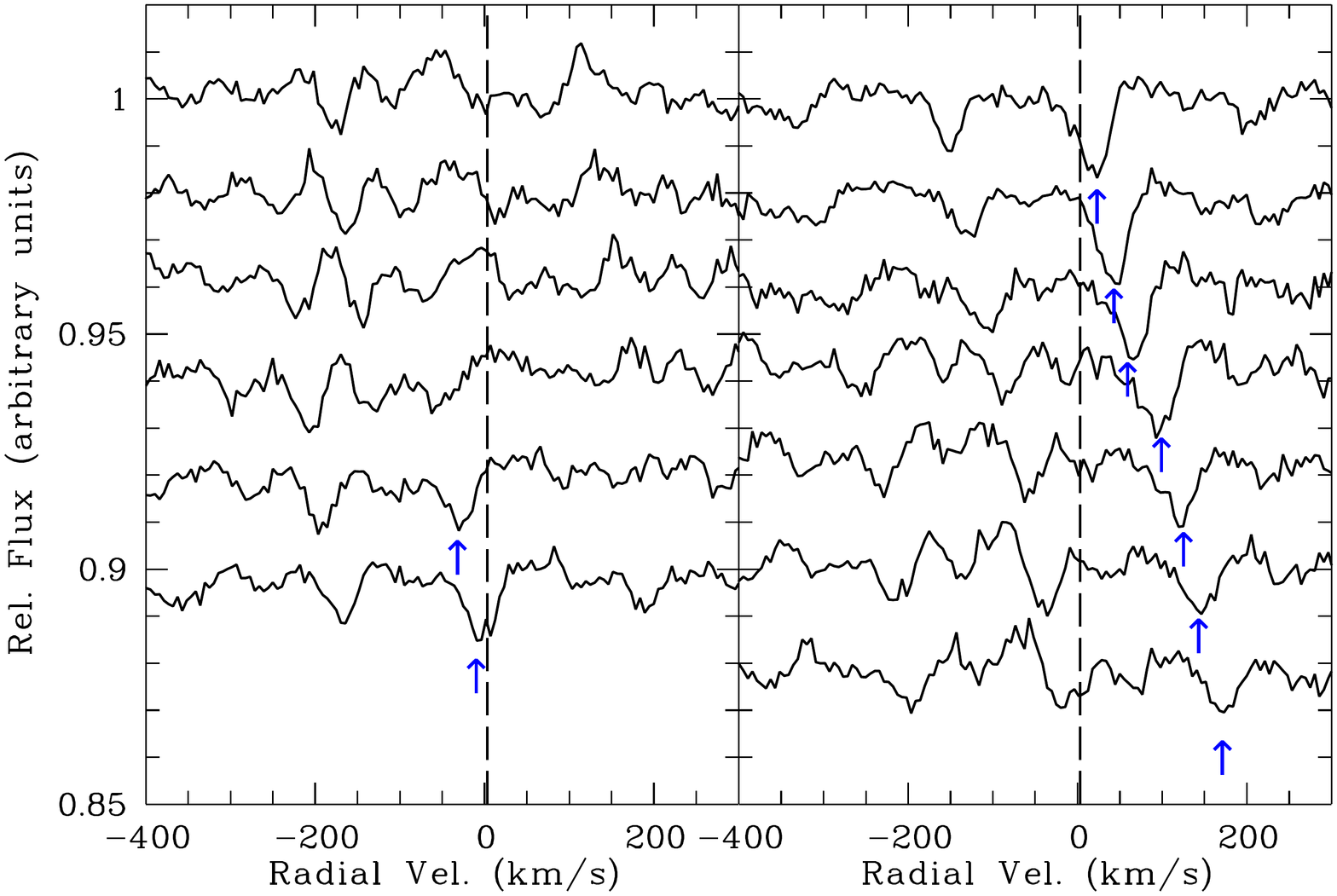}}
\caption{Ca {\sc  i} 4226 \AA  ~and Mg  {\sc ii} 4481  \AA ~absorption
  lines as observed in the 13 consecutive HARPS  spectra taken
  during  the night  2019-01-29/30.  Exposure  time was  15 minutes  per
  spectrum.  Dates and UTs of the spectra are given. Pixel  resolution is 5.5
  km/s.   The top  panels show  the observed  line profiles  while the
  bottom panels  show the  residuals once the  median spectrum  of the
  night has been removed. The median spectrum of the night is plotted in
  red. The blue arrows mark some examples of bumps (in the Ca {\sc
    i}  panels) and  dumps (in  the Mg  {\sc ii}  panels).  The  black
  dashed line in all panels corresponds to a radial velocity of 3 km/s
  (see  text  in Subsection  \ref{sec:variabilityofcsabsorptions}  and
  Fig. \ref{cs:velocities_median})}
\label{2019_cai_mgii} 
\end{figure*}

As  already  mentioned,  the  median   spectrum  of  $\phi$  Leo  shows
photospheric   lines  typical   of  a   late  A-type   spectrum  (Fig.
\ref{photospheric}), but individual spectra distinctly exhibit that at
least some of  those lines (e.g.  Ca  {\sc i} 4226.7 \AA,  Ti {\sc ii}
4468.5 and 4501.3 \AA,  Mg {\sc ii} 4481 \AA, Fe  {\sc ii} 4508.3 \AA)
vary, with the aforementioned dumps  and bumps appearing superimposed on 
the line profiles. Fig. \ref{201603_cai_mgii} displays  the absorption
profiles of the Ca {\sc i} 4226  \AA ~and Mg {\sc ii} 4481 \AA ~lines,
as observed  in the three  spectra taken  with FEROS during  the night
2016-03-28/29.   Very clear  extra  absorptions close  to  the core  are
observed in comparison  to the median  spectrum of the whole  March 2016 period;
the amplitude of variations in the  equivalent widths of both lines being
small, in  the range of  2-3 times the estimated  uncertainties (Table
\ref{tab:EWs_mg_ca}).

Time series spectra taken during  the May 2016, 2017, and 2019 observing
runs allow  us to  better characterise the  changes in  the absorption
lines, and  to trace their  temporal evolution with a  time resolution
between 4  and 15 minutes  during a time  interval from $\sim$3  up to
$\sim$9  hours  (during some  nights  of  March-April 2017 HERMES and  
FEROS spectra were consecutively collected for  a time interval of $\sim$9.5
hours,  even   some  spectra  were  taken   simultaneously  from  both
telescopes).  Fig.   \ref{2019_cai_mgii} shows  the profile of  the Ca
{\sc i} 4226 \AA ~and Mg {\sc ii} 4481 \AA ~lines corresponding to the
13 consecutive  spectra obtained with  HARPS-N in January  2019 during
$\sim$3 hours of observations -integration time of each spectrum was
15 minutes.  Several dumps and bumps propagating across the profile at
different radial  velocities can be  distinguished; in both  lines the
features vary  their radial velocity  at a rate  of $\sim$100  km/s per
hour (to  guide the  eye some of  them are marked  with arrows  in the
figure). Fig. \ref{2019_cai_mgii} shows the median of the HARPS-N spectra
superimposed on each observed individual spectrum as well as the residuals 
once the median  spectrum has  been  removed. The   features  (dumps/bumps) 
propagate  towards   redder  velocities  first  increasing  their strength, 
achieving a maximum at a velocity of $\sim$3 km/s, i.e., close to the apparent emission detected in CS features 
(see  details in Subsection \ref{sec:variabilityofcsabsorptions} and Fig.
\ref{cs:velocities_median}), and  then
decreasing their strength, seemingly  lasting from $\sim$1 hour to
$\sim$3  hours; their  persistence from  profile to  profile suggesting
that  they are  real and  not artifacts  or noise.   Similar features,
varying their strength and sharpness,  and moving from blue-shifted to
more red-shifted  velocities superimposed on the  median line profiles
are  seen in  other  nights.  We note  that  the  time span,  velocity
interval and maximum strength are irregular, and that the behaviour of
the Ca {\sc  i} and Mg {\sc  ii} lines is not quite  coincident, or in
other   words,  the   variations  do   not  seem   to  be   in  phase.
The fact  that they  do not correlate  might occur  if the
  lines  originate  at different  layers  in  the stellar  photosphere.
We also note that  dumps/bumps are not always clearly
present in many spectra. Appendix  A presents some further spectra
of the  Ca {\sc i}  4266 \AA ~and Mg  {\sc ii} 4481  \AA ~photospheric
lines.

\subsection{Circumstellar absorptions}
\label{sec:variabilityofcsabsorptions}
\begin{figure*}[!ht] 
\centering
\scalebox{0.6}{\includegraphics[angle=-90]{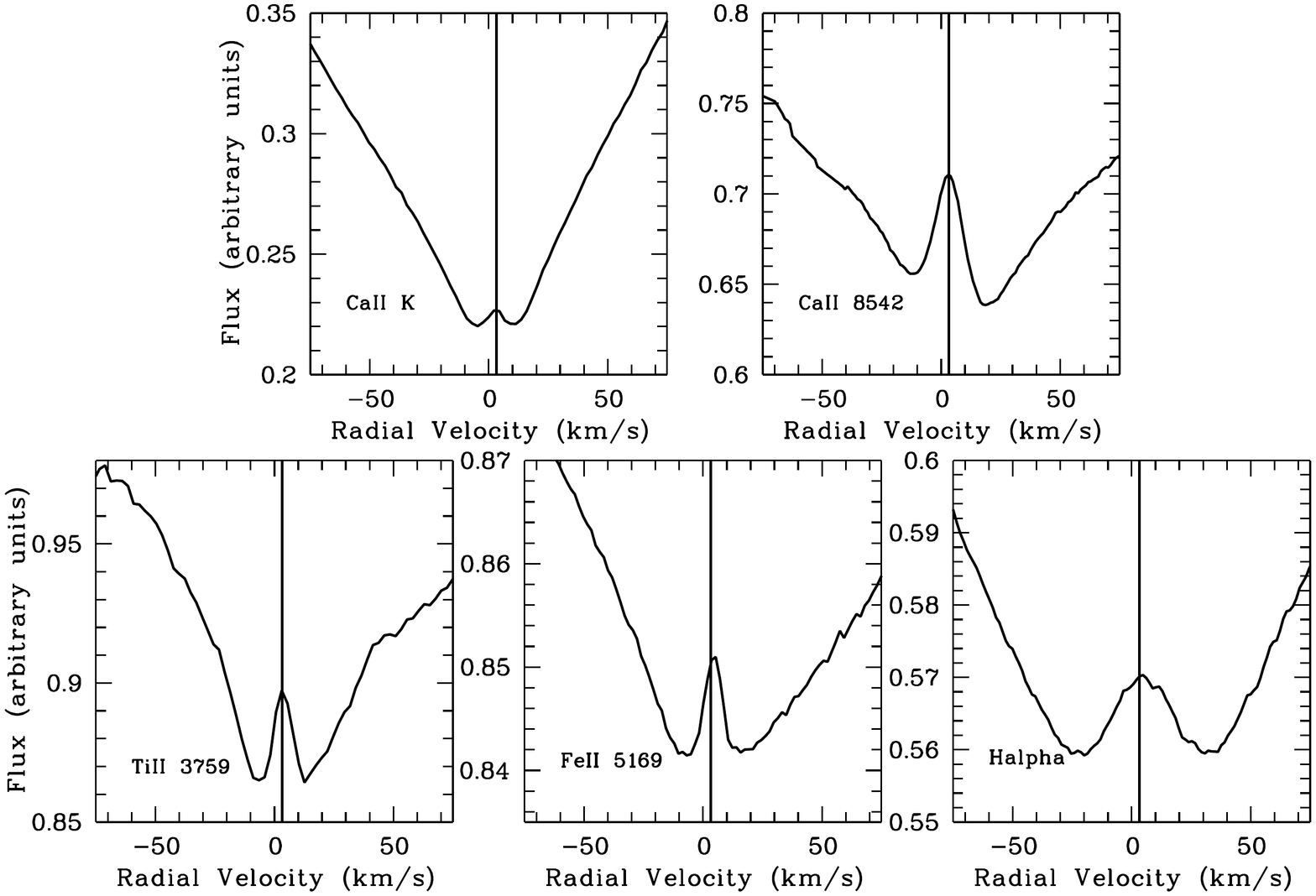}}
\caption{Line  profiles  in  velocity units  of  metallic  absorptions
  showing  a clear  non-photospheric contribution.  H$\alpha$ is  also
  plotted.  The  profiles correspond to  the median of all  $\phi$ Leo
  HERMES and FEROS spectra. The  vertical line corresponds to a radial
  velocity of 3 km/s. In the case of Ca {\sc ii} K line the profile corresponds to 
  the ``pure'' CS contribution, i.e, the photospheric absorption has been removed (see text).}
\label{cs:velocities_median} 
\end{figure*}

\begin{figure*}[!ht] 
\centering
\scalebox{0.3}{\includegraphics[angle=-90]{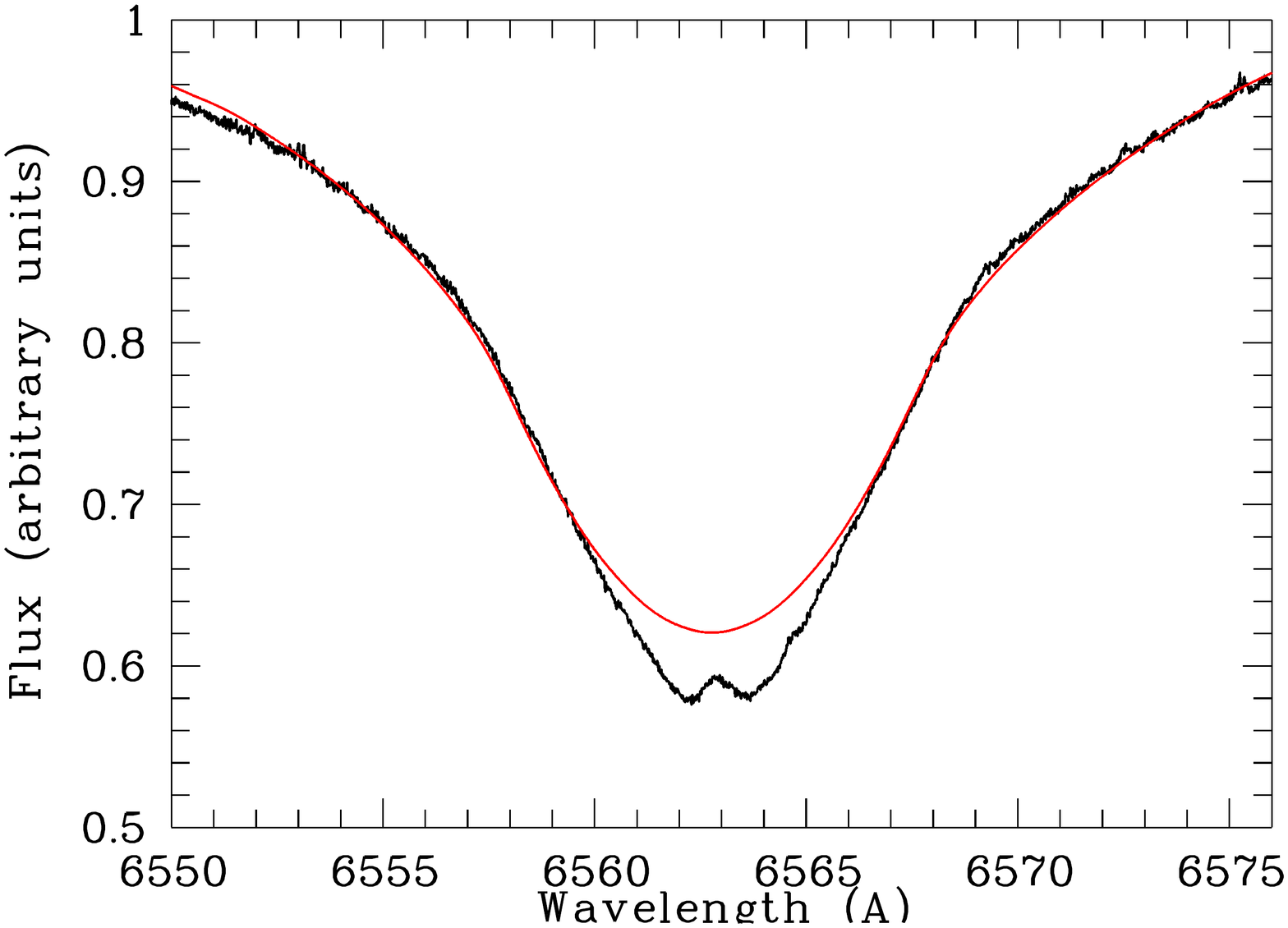}}
\scalebox{0.3}{\includegraphics[angle=-90]{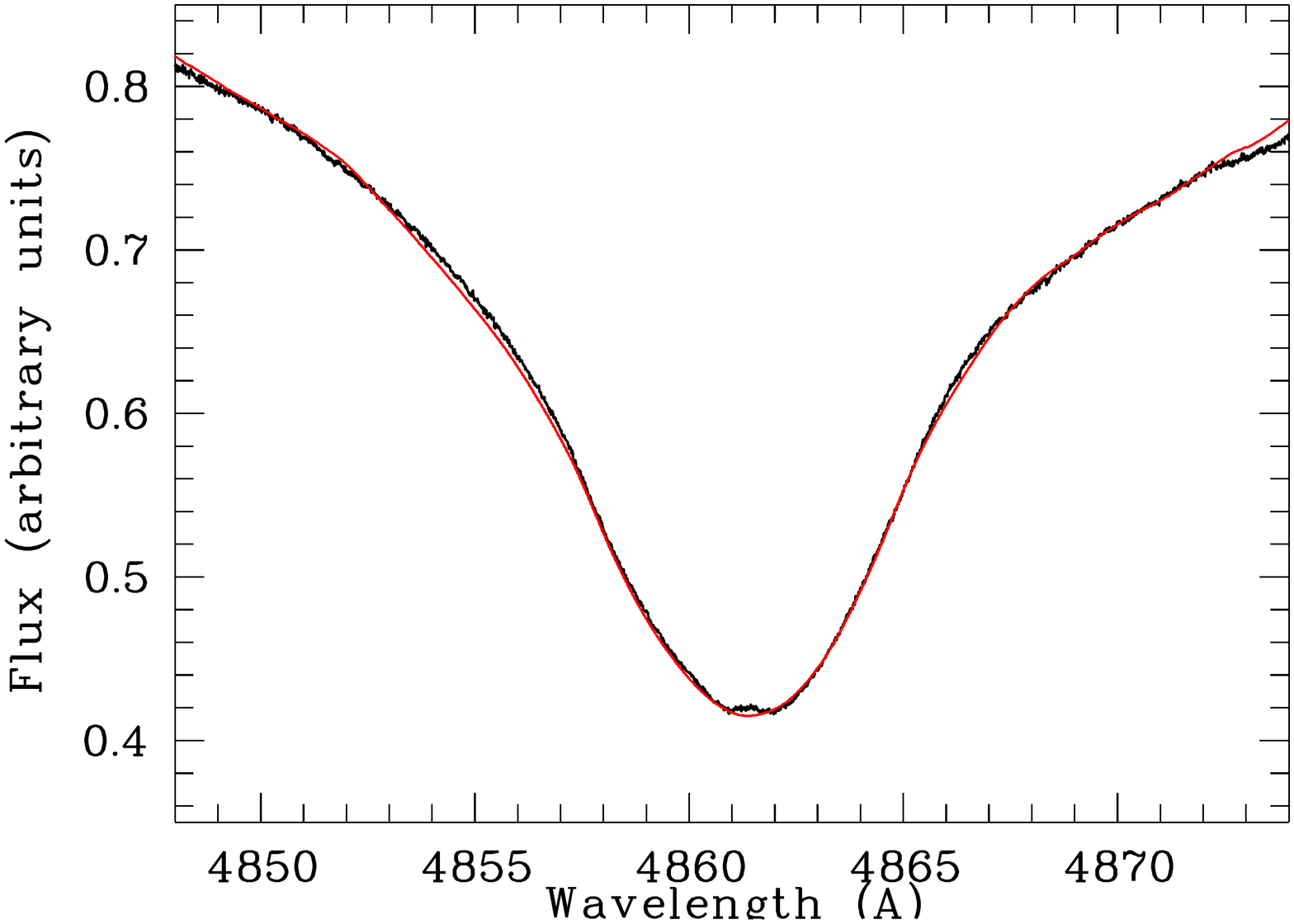}}
\caption{Median profiles of the H$\alpha$ and H$\beta$ lines taken with HARPS 
with the Kurucz model superimposed.}
\label{halpha_hbeta_excess} 
\end{figure*}

\begin{figure*}[!tb] 
 \centering
 \scalebox{0.5}{\includegraphics[angle=-90]{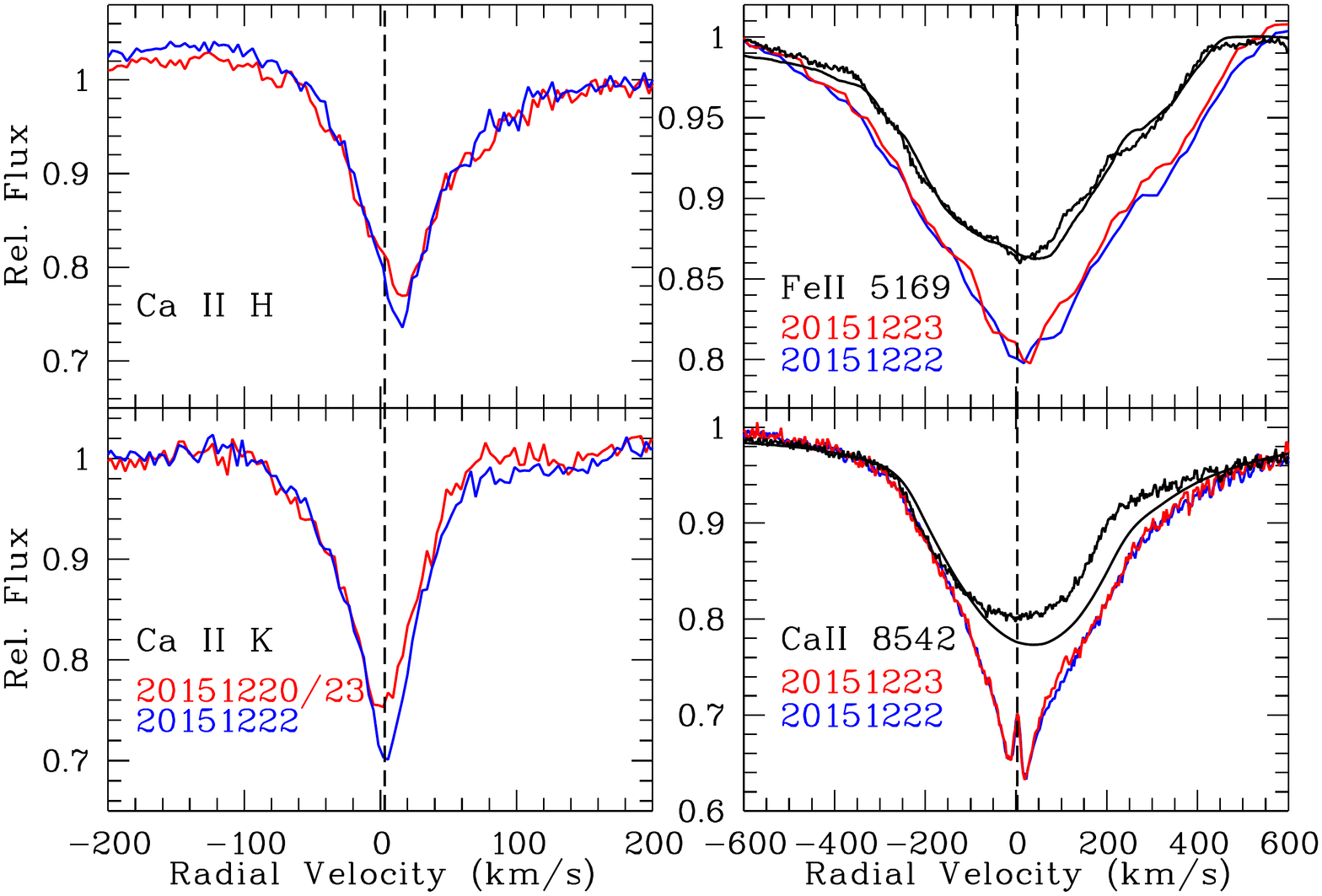}}
\caption{Left:  Circumstellar  Ca  {\sc  ii} H  (up)  and  K  (bottom)
  features  of  the  HERMES  spectra  taken  in  December  2015.   The
  photospheric line has  been removed as estimated from Kurucz 
  synthetic models (this is  the procedure followed for all Ca {\sc ii} 
  H\&K plots, see Sect. \ref{sec:variabilityofcsabsorptions}). The spectra 
  have been rebinned to 3 pixels (size 3.56 km/s). Right (bottom):  Ca {\sc ii} 
  8542 \AA.  The spectra  have  been  rebinned  to  a resolution  of  3.29  
  km/s  per pixel. Right (up)  Fe {\sc ii} 5169 \AA ~line;  the pixel resolution
  in this  case is  8.16 km/s.  The  photospheric contribution  is not
  removed in the right panels (this  is the procedure for all plots of
  these  lines).  Dates  are  colour-coded.  The  vertical dashed  line
  corresponds to a  radial velocity of 3 km/ (this  line is plotted in
  all plots).   With the  aim of  comparison and  to visualize  the CS
  contribution, the  right panels also show  in black the Fe  {\sc ii}
  5169 \AA ~and Ca {\sc ii} 8542  lines of the $\phi$ Leo Kurucz model
  and the observed spectrum of the star $\alpha$ Aql.}
\label{mercator_201512} 
\end{figure*}

Prominent non-photospheric shell-like features are seen in the Ca {\sc
  ii} H\&K lines, the Ca {\sc ii}  triplet at 8498, 8542 and 8662 \AA,
the Ti {\sc ii} lines at 3685,  3759 and 3761 \AA, the  Fe {\sc ii}
 triplet at 4923.9, 5018.4 and  5169 \AA, and the  Balmer lines
H$\alpha$ and H$\beta$. Other shell lines as those  observed in other
shell  stars,  see  e.g.   the  spectrum of  HR  7731  shown  in  Fig.
\ref{nonphotospheric}, are either hardly distinguishable or undetected.
Fig.  \ref{cs:velocities_median} shows  the median HERMES and FEROS
spectrum in velocity space of some of those lines - in the case of the
triplet lines the  most representative one is shown.  These lines show
common and distinct behaviours as explained in the following:

\noindent  i) The  median  spectrum of  the non-photospheric  metallic
features   shows   a   prominent   emission   at   the   line   centre
(Fig. \ref{cs:velocities_median}).  The emission peak is  located at a
velocity of $\sim$3 km/s.  The emission at the  core is also present
in the H$\alpha$  line and very weak in H$\beta$.  We  note that  the
emission in the Ca {\sc  ii} H\&K lines is not apparent in the 2015 spectra 
(see Fig. \ref{mercator_201512}) but it clearly appeared in the 2017 and 2019
spectra, as well as in the 2016 spectra (see Fig. \ref{feros_201603_caiihk}).
In both Fig. \ref{mercator_201512}  and \ref{feros_201603_caiihk} the photospheric line contribution has been removed - the procedure we have followed to visualize the 
Ca {\sc ii} H\&K CS features is described below after item v). 

\noindent ii) Manifest variability is observed in the Ca {\sc ii} H\&K
and Fe {\sc ii}  triplet lines, as well as a less pronounced variation in the  
Ti {\sc ii} lines.  Changes  are observed on any time scale  ranging from
minutes to days or  observing runs, as perceived when  comparing spectra taken
during the same  night, or the median spectra of  the different nights
or  observing  periods.  The  variations  do  not show  a  discernible
temporal pattern, and there is  no correlation between the observed variability
of the different metallic ions.

\noindent iii)  As in the metallic lines, an absorption excess  is clearly 
visible at the core of H$\alpha$ (Fig. \ref{halpha_hbeta_excess}). H$\alpha$ 
and  H$\beta$ remained without  any noticeable
changes in the spectra taken with  HERMES and FEROS from 2015 December
to 2017 April, but H$\alpha$ displayed a clear variability in the spectra
of 2019-01-29 taken with HARPS-N.   These variations  do not
exactly correlate  with those simultaneously  observed in the  Ca {\sc
  ii} and Fe {\sc ii} lines.

\noindent iv) The  Ca {\sc ii} triplet lines do  not show any apparent
change when spectra are compared in any time scale. We note, however, that the HARPS-N 
spectra do not cover these lines; thus, we do not know if  these 
lines varied in 2019-01-29 as H$\alpha$ did.

\noindent v) In  all CS lines, including H$\alpha$, the variability shows up as 
changes in the relative absorption  strength at both blue and red  minima of the
central emission, but the strength of those minima in the median spectrum is
practically indistinguishable (Fig.  \ref{cs:velocities_median}). In  the 
case  of the  constant Ca {\sc ii}  triplet lines, the blue side is always weaker
than the red side. Further, for each line the minima appear  at nearly  the same
radial velocities, 

Spectra of some  observing runs illustrating the mentioned points above are 
shown in the next subsections. Beforehand, we point out that in the case of the Ca
{\sc ii} H\&K lines the plots show the CS features once the photospheric 
contribution has been removed. This is carried out by dividing the observed profiles
by synthetic ones estimated using the  Kurucz synthetic spectrum computed with  
the T$_{\rm eff}$, $\log  g$ and $v \sin i$ values of Section \ref{hr4368} 
as it  provides a very reasonable  fit  -see Fig.  1 in \cite{eiroa16}. We are  
aware  that due  to  the  high  rotational velocity, the oblateness  of $\phi$ Leo, 
as  mentioned in Section \ref{hr4368}, could be significant, and therefore  
due to gravity darkening  a  gradient  from  equator  to  poles  
in at least T$_{\rm eff}$, $\log g$  must be present. Nonetheless, we
are  confident that  our  estimates provide reasonable mean values
since our Kurucz  model for $\alpha$ Aql gives a very good fit 
\citep[see Fig. 1 of ][]{eiroa16}, this  star also being affected by gravity
darkening due to its relevant oblateness \citep[e.g.][]{bouchaud20}. 

\subsubsection{HERMES December 2015}

The Ca  {\sc ii} H\&K  lines of December  20 and 23  are similar
showing  a  broad  non-photospheric  absorption  in  addition  to  the
photospheric line  - see e.g. Fig.   1 by \cite{eiroa16}; at  the same
time,  the  four spectra  of  December 22  remained  constant,  and  the
absorption at the core was deeper than in the other two nights.  Based
on  this  result,  and  on   the  behaviour  of  other  spectra obtained 
in 2016, \cite{eiroa16}  assumed  that  the  spectra  of  December  20  
and  23 represented  a stable  absorption formed  by the  contribution of  the
stellar photosphere and  a CS gas disk, which was  taken as a template
to analyze other contributions/variations to the CS gas.  However, the
detection of the  central 3 km/s emission in the  spectra of 2017 and
2019,  which at  least  was  not clearly  present  in  2015,
provides  a  new  perspective,   indicating  that  the  assumption in 
our previous work \citep{eiroa16} was not  appropriate. We now note that 
the December 22 spectra  are among the  deepest absorption  of all
spectra taken in  different observing runs,  and that  the December 20  and 23
spectra have depths similar to many other spectra. Also, the profile of the CS Ca
{\sc ii}  H\&K absorption in  December 2015  suggests that the 3 km/s
emission at  the core was hidden  or not present.   
Thus, our analysis is  now carried
out considering  the whole  absorption in  excess of  the photospheric
line given by our  Kurucz  model.

Fig.  \ref{mercator_201512}  (left) shows  the nightly average  of the
non-photospheric Ca {\sc ii} H\&K  features once the photospheric line
has been removed.   It can be observed that both  Ca {\sc ii} H\&K features are
deeper in December 22 than in  December 20 and 23.  The estimated equivalent widths of the K/H features are $\sim$227 m\AA/$\sim$234 m\AA, 
and $\sim$249 m\AA/$\sim$245 m\AA, for the spectra December 20/23 and December22, respectively;  these figures suggest that  the Ca  {\sc  ii} features  are  
saturated. The  peak  of the  K absorption is centered at the velocity 
of the 3 km/s central emission, although as pointed out above such emission is not present 
in these spectra, and the H line appears  at a  velocity  of $\sim$15  km/s.  
Such  a  shift to  more redshifted   velocities is observed in other shell
stars \citep{rebollido20},  and  is  likely  due to  the  inference  of  the
H$\epsilon$ line.  Fig.  \ref{mercator_201512}  (right) also shows the
Fe {\sc ii} 5169 \AA ~and the Ca  {\sc ii} 8542 \AA ~ lines, where one
can appreciate  that Ca {\sc ii}  8542 \AA ~does not  vary and clearly
shows the central 3 km/s emission,  while the Fe {\sc ii} line neither
appreciably varies (i.e. it differs in  behaviour from the Ca {\sc ii}
H\&K lines) nor the central emission is present.  The Balmer H$\alpha$
and H$\beta$ lines -not shown- are constant and present  the 3.0 km/s
central emission.

\subsubsection{FEROS March 2016}

\begin{figure*}[!ht]
  \centering
\scalebox{0.285}{\includegraphics{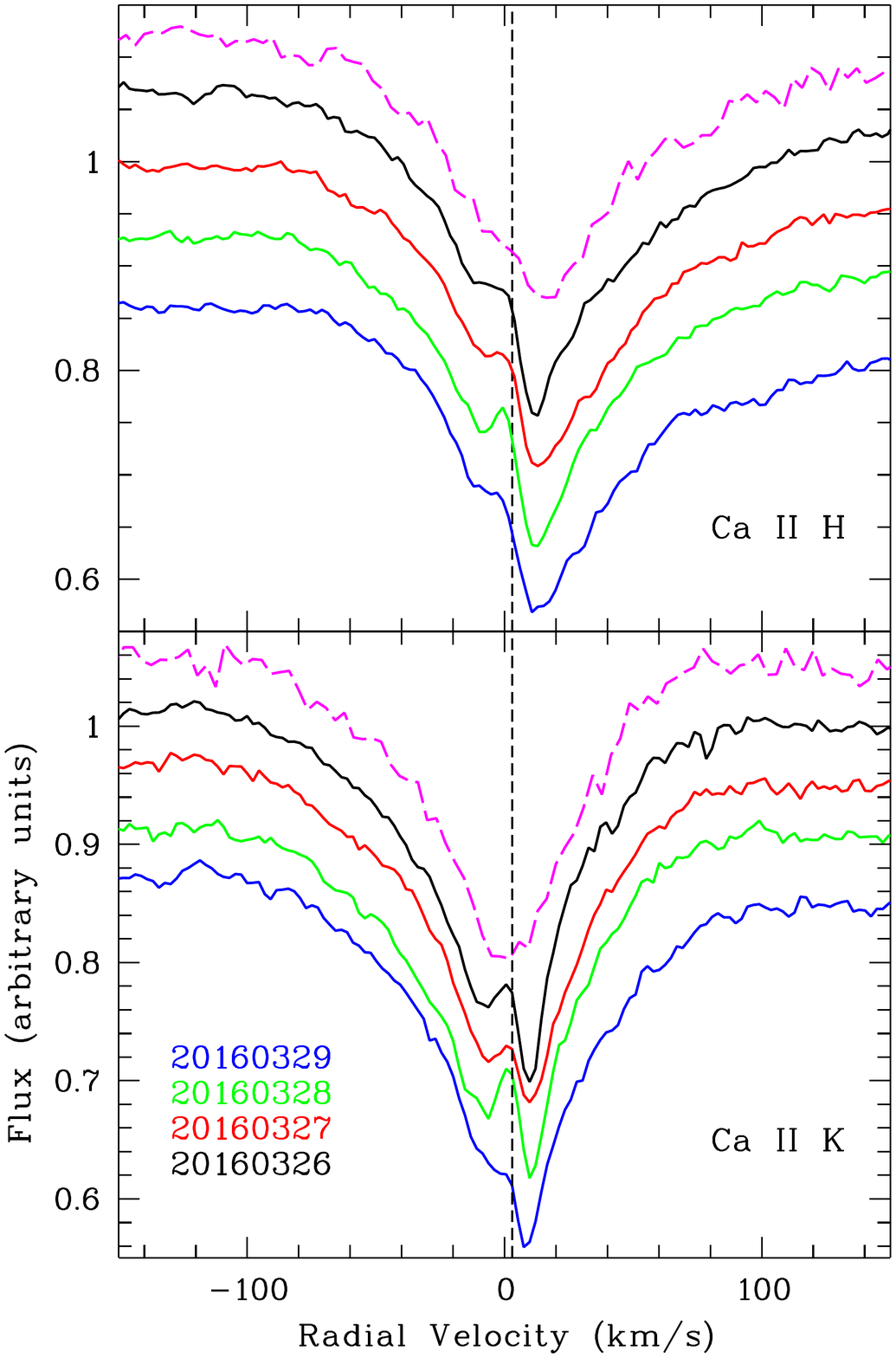}} 
\scalebox{0.285}{\includegraphics{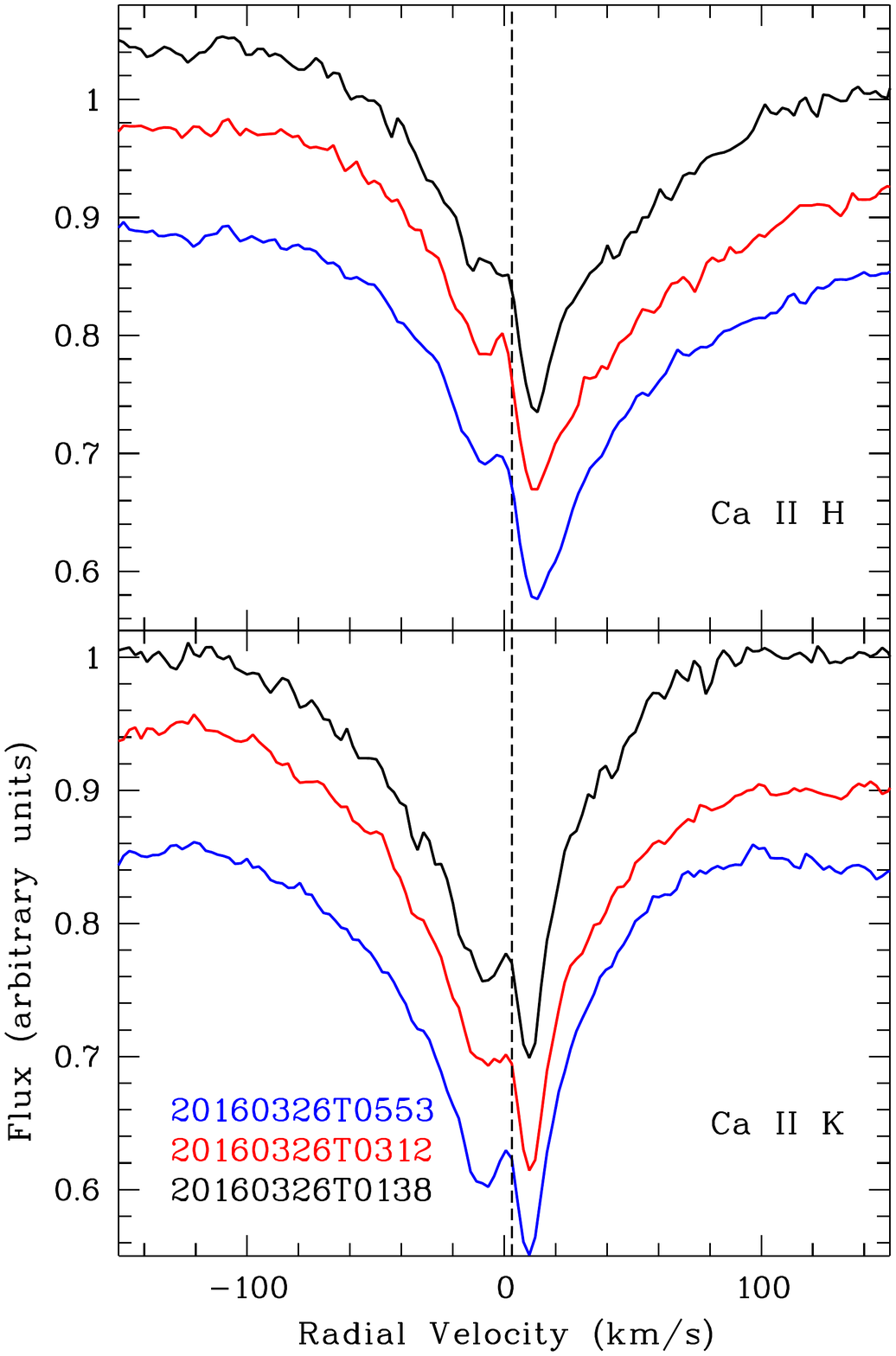}}  
\scalebox{0.285}{\includegraphics{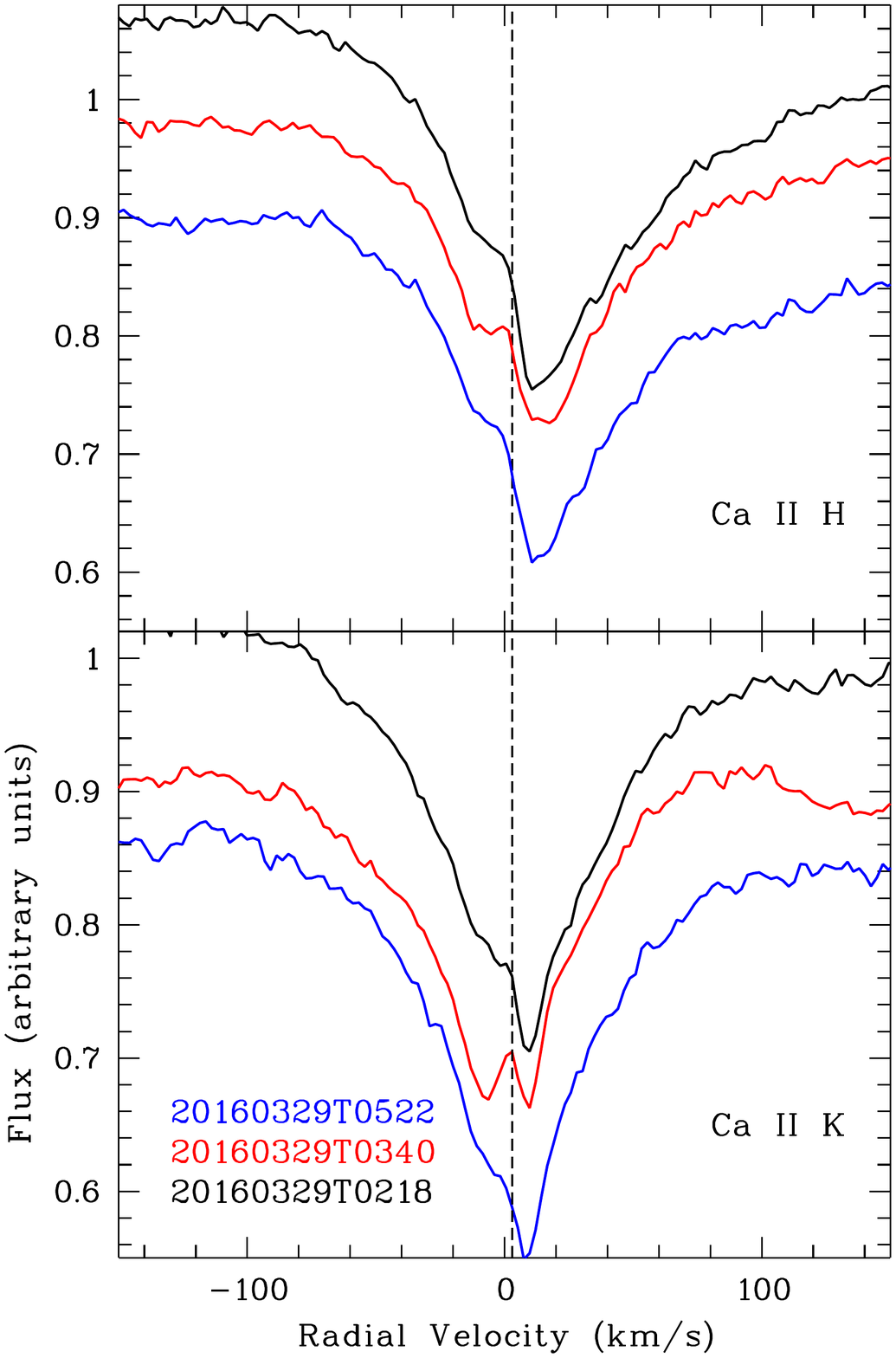}}
\caption{Ca  {\sc  ii} H\&K  spectra  taken  with FEROS  during  March
  2016. The  left panel plots  the median of  all 4 nights,  while the
  middle and right panels shows the observed features in two different
  nights. UTs  are indicated  in the labels.  With the aim of comparison the
  2015-12-20/23 spectrum is shown in  the left panel (dashed lines).}
\label{feros_201603_caiihk} 
\end{figure*}

Fig. \ref{feros_201603_caiihk} shows the Ca  {\sc ii} H\&K CS features
as observed in this period once the photospheric contribution has been
removed. Variability can be observed on  a nightly basis as well as in
the median  of each of the  four observed nights; we  see that, unlike
the procedure followed  by \cite{eiroa16}, the 3  km/s emission feature
is  distinctly distinguishable  in many  of both  individual and  median
spectra. With the  aim of comparison, the feature as  observed in 
2015-12-20/23  is also  plotted in  the figure.  In  this way  it can
easily  be noticed  that the  variable  absorption at  both blue-  and
red-shitfted sides  of the  central emission can  be misled  with FEBs
events  similar   to  those   of  $\beta$  Pic,   if  as   assumed  by
\cite{eiroa16}  the  2015-12-20/23  feature is  taken  as  a  stable
template.   Besides this,  we point  out  that the  variations in  the
strength  of the  CS absorption  at  both blue  and red  sides of  the
central  emission  is recursive  in  other  observing runs,  that  the
changes do not  follow an identifiable temporal pattern,  and that the
velocity of  the blue and  red deepest absorptions is  always similar,
$\sim-10/-6$ km/s and $\sim10/20$ km/s, respectively.

\begin{figure*}[!ht]
  \centering
\scalebox{0.2045}{\includegraphics[angle=-90]{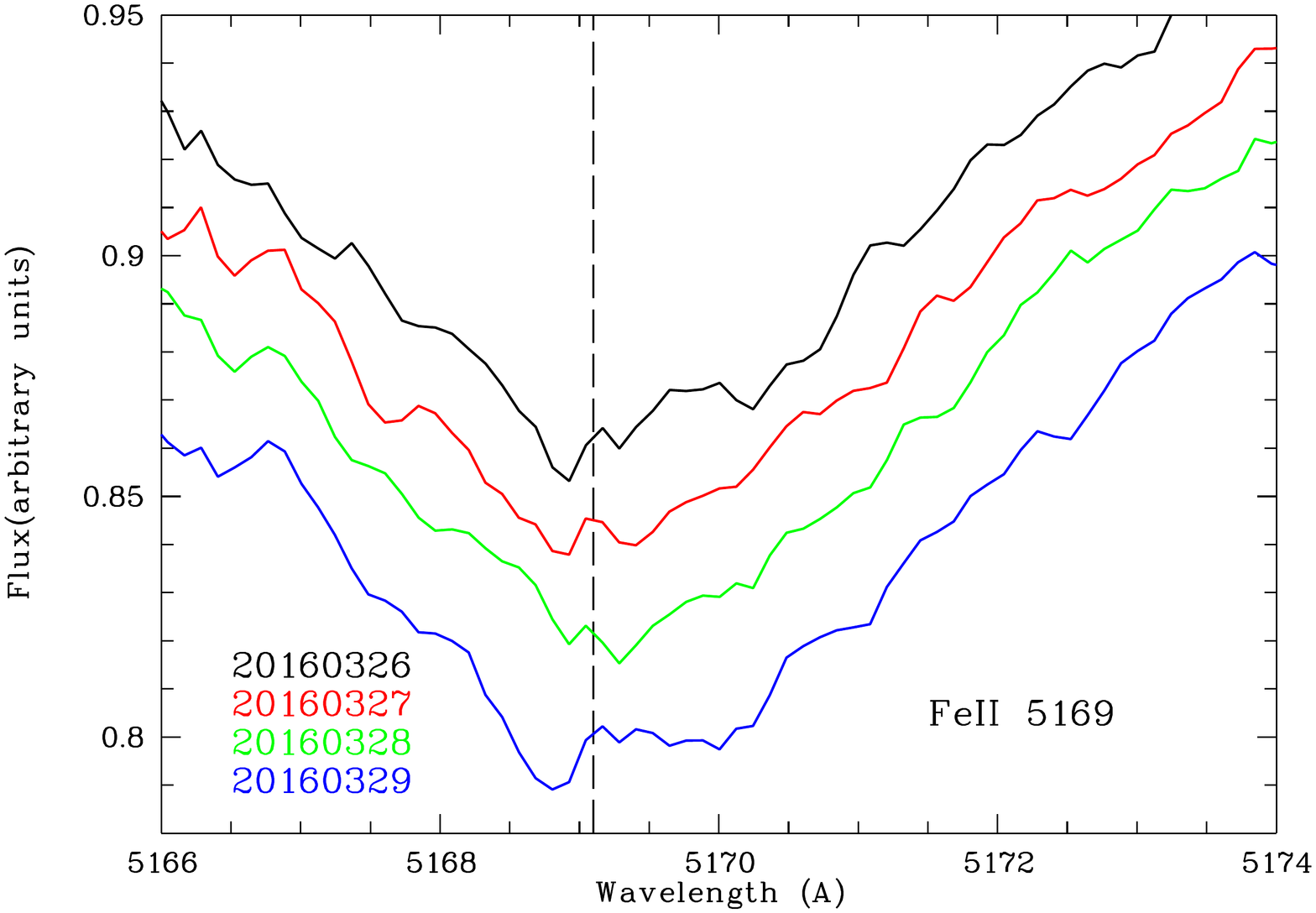}}
\scalebox{0.2045}{\includegraphics[angle=-90]{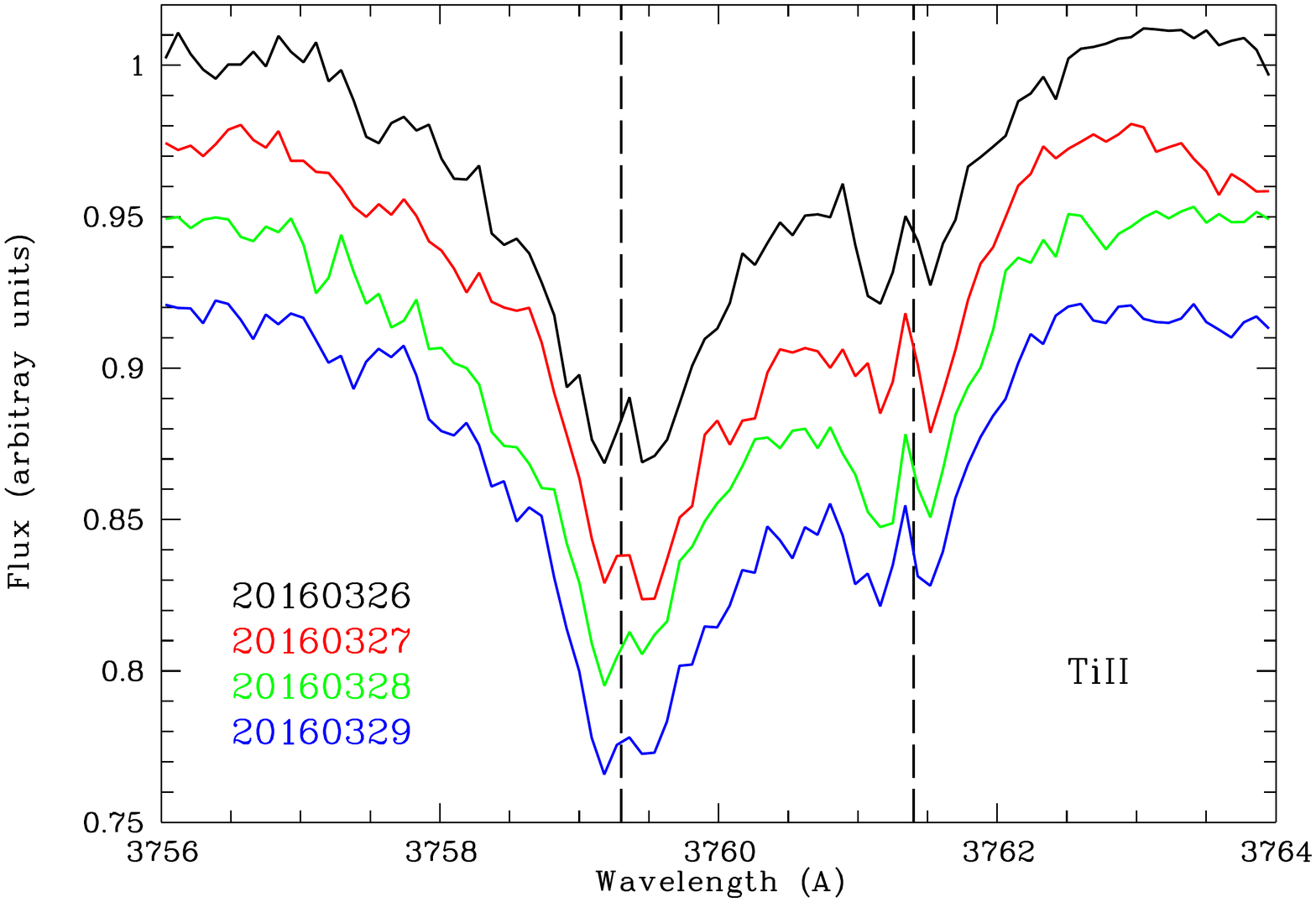}}
\scalebox{0.2045}{\includegraphics[angle=-90]{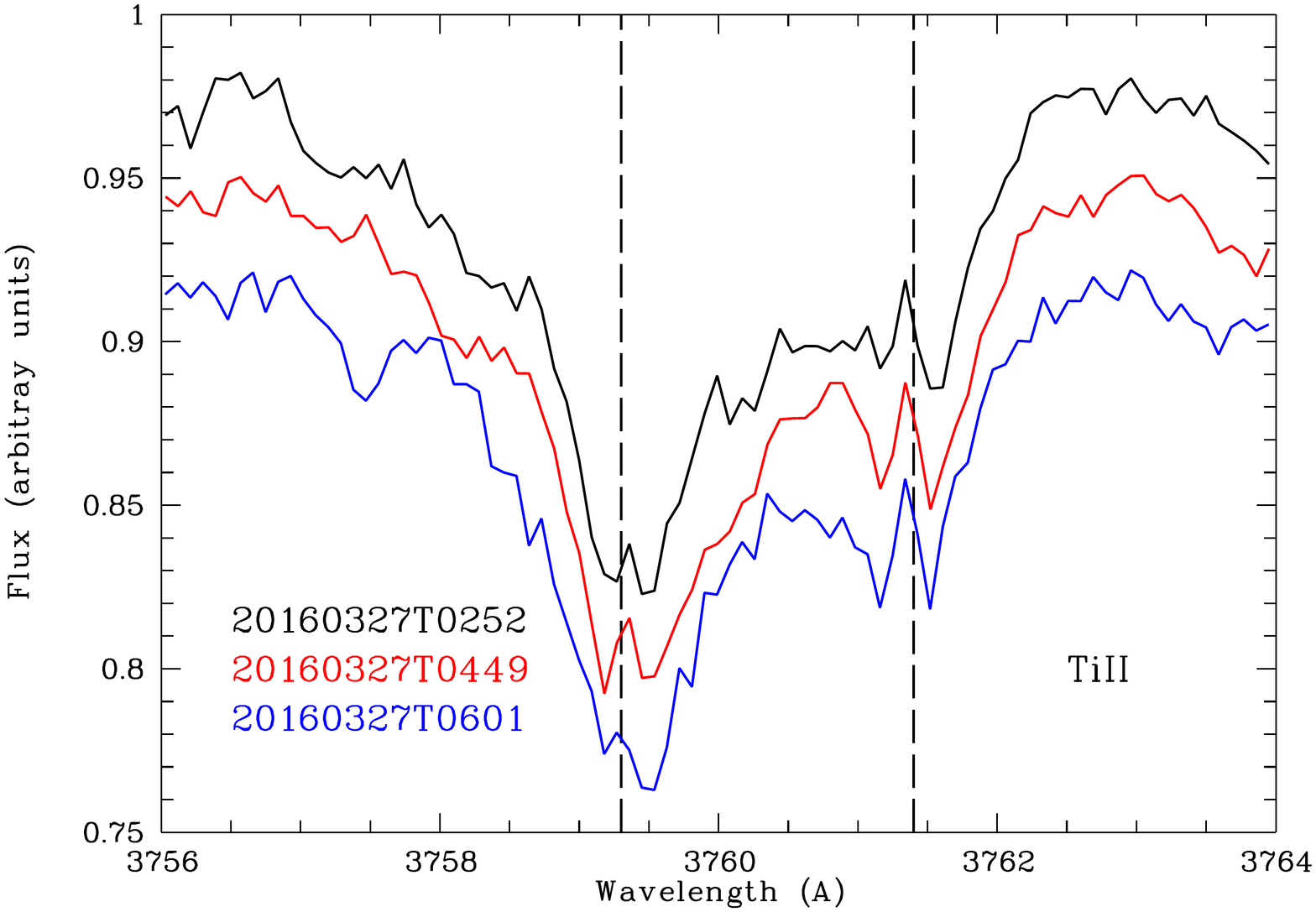}}
\caption{Left:Fe {\sc  ii} 5169 \AA  ~median daily spectra  taken with
  FEROS during  March 2016.   Middle: median daily  spectra of  the Ti
  {\sc ii} lines 3759  and 3761 \AA. Right: same Ti  {\sc ii} lines as
  observed  in  the  night 2016-03-26/27.  Resolution  is  1.2
  \AA/pixel. UTs are indicated in the  labels. In all plots the dashed
  line corresponds to a 3 km/s velocity.}
\label{feros_201603_5169_ti} 
\end{figure*}

Fig.\ref{feros_201603_5169_ti} plots  the Fe {\sc  ii} 5169 \AA  ~ and
the Ti {\sc  ii} 3759 and 3761 \AA ~lines. In the case of  the Fe {\sc ii}
line  we have chosen to  show the median spectra  of each night,
while in the case of the Ti  {\sc ii} lines the median spectra as well
as the individual spectra of the  night 26/27 March are plotted. The 3
km/s emission is seen in  all cases, being particularly conspicuous in
all spectra of both  Ti {\sc ii} lines. As in the case  of the Ca {\sc
  ii} K line changes in the strength of the lines at both sides of the
central  emission are  observed,  but without  any clear  correlation
among the variations in the different lines. Just to mention,  there are
no identifiable changes in  the Balmer and the Ca {\sc ii} triplet lines.

\subsubsection{HARPS-N January 2019}
\label{harps}

\begin{figure*}[!ht]
  \centering
  \scalebox{0.2896}{\includegraphics{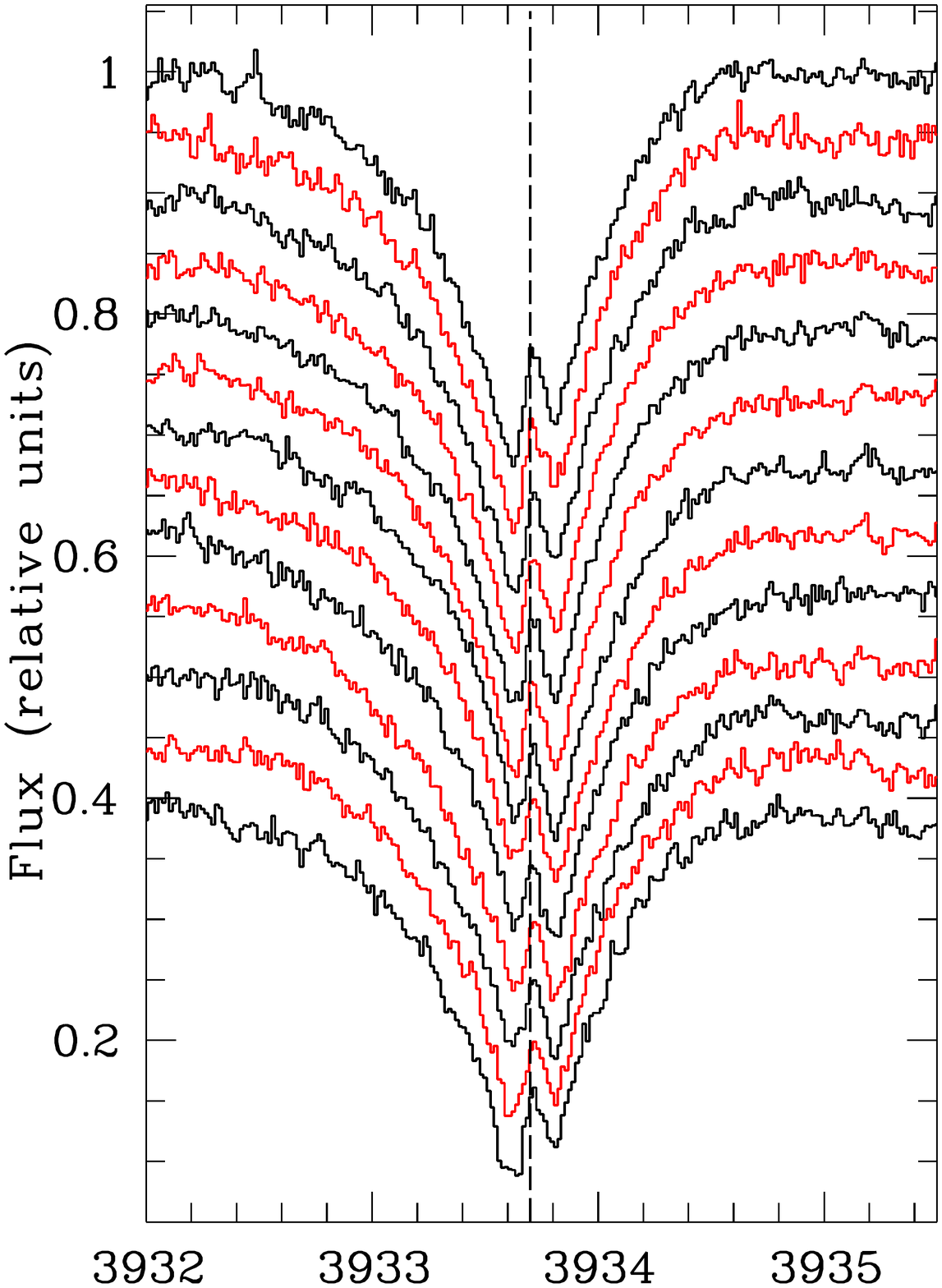}} 
  \scalebox{0.2896}{\includegraphics{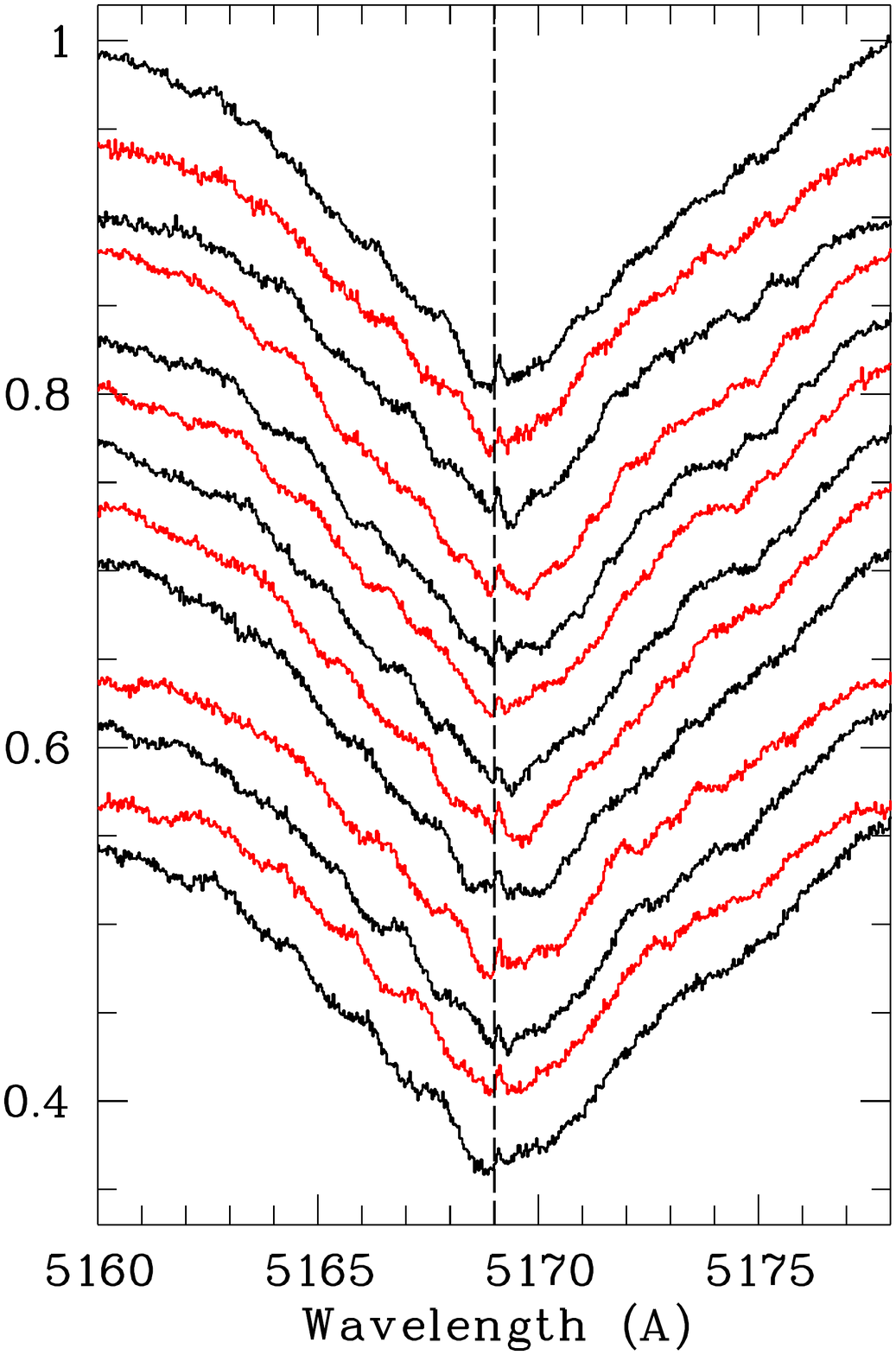}}
  \scalebox{0.2896}{\includegraphics{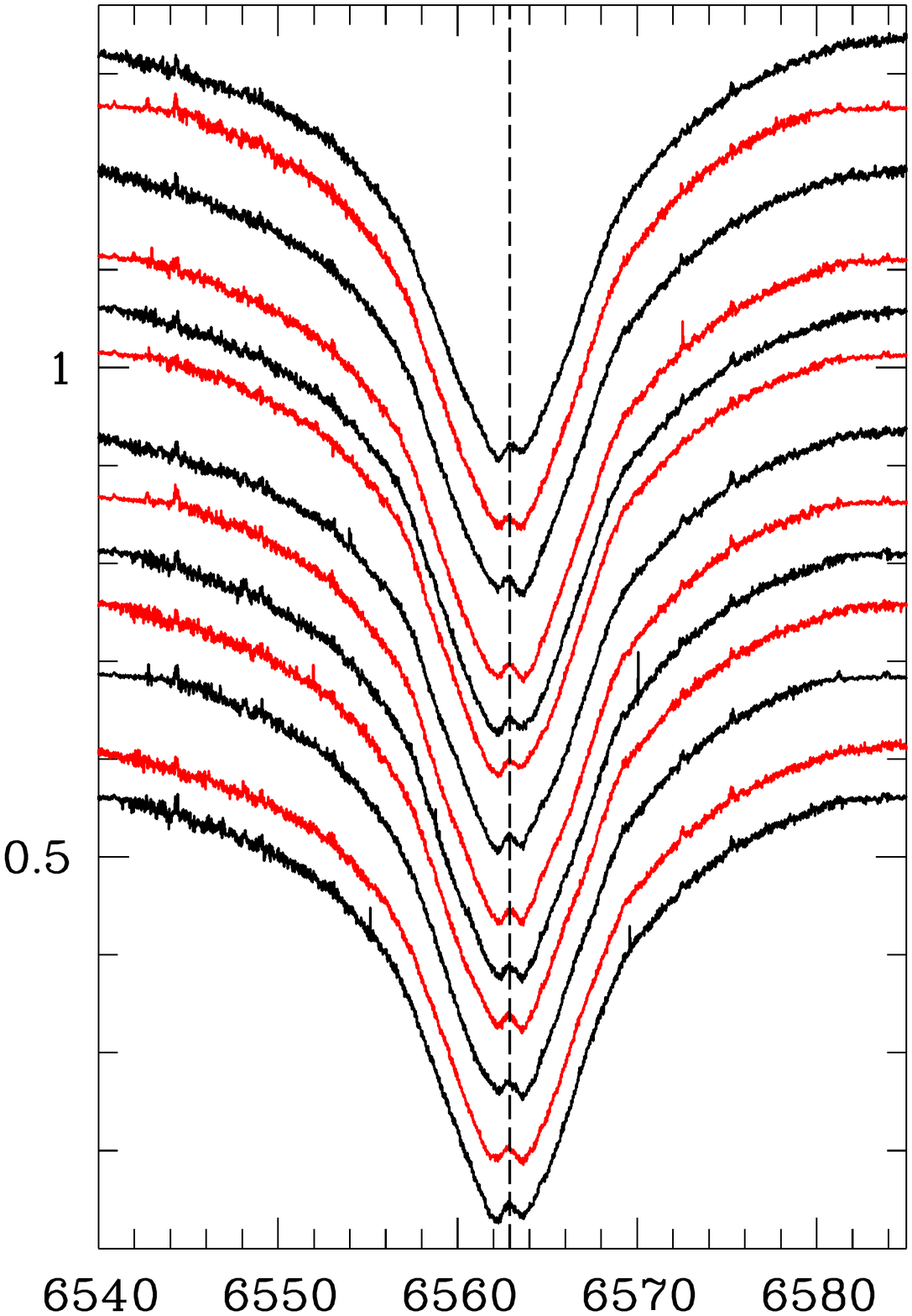}}
\caption{Ca {\sc ii} K (left panel, resolution 0.016 \AA/pixel), Fe {\sc ii} 5169 \AA ~(middle, resolution 0.03 \AA/pixel)), H$\alpha$ (right, resolution 0.01  \AA/pixel).  Spectra were taken with HARPS-N during the night 2019-01-29/30; each spectrum is  15 minutes apart starting 01h01m UT (first spectrum at the top), and finishing 04h07m UT (last spectrum at the bottom). The dashed line corresponds to a velocity of 3 km/s.}
\label{harps_201901_cs} 
\end{figure*}

Fig. \ref{harps_201901_cs} shows the Ca {\sc ii} K, Fe {\sc ii} 5169 \AA, and
H$\alpha$  as observed in the 13 consecutive spectra taken with HARPS during the
night of 2019-01-29/30, with an exposure time of 15 minutes each, 
the first at the top starting at 2019-01-30 01h01m UT. 
The photospheric contribution has been removed
in the case of the Ca {\sc ii} lines. The central emission is clearly visible
in all Ca {\sc ii} and  H$\alpha$ spectra, but it is not as obvious  in the 
case of some Fe {\sc ii} line profiles. Variability is seen at both sides of 
the central emission among the individual consecutive spectra in all three 
lines; again that variability does not follow a regular temporal pattern, 
neither there is a correspondence in the variability of the different lines. 
We note, as in previous spectra of other observing runs, that the blue and 
red absorption peaks occurs at practically the same velocity without any 
evolution in velocity. 

\section{Discussion}

\subsection{The $\delta$ Scuti photosphere}

The temporal  variation of the  photospheric lines and the  $\delta$ 
Scuti {\em TESS}   photometric  variability   \citep{balona20} are alike.  
As   stated  in Sect.  \ref{hr4368}, $\phi$  Leo is  most likely  
seen edge-on; thus, assuming  a stellar  radius  of  3.2 $R_\odot$  
and  230  km/s as  the equatorial  velocity,  the  rotation period  
would  be  $P\!\approx$17 hours,  i.e., seemingly larger  than  
the apparent  temporal  duration  of  the features crossing the 
photospheric lines. While the uncertainty of the
assumed stellar parameters is high since this estimate does not
take into  account the oblateness of  the star - and  consequently the
expected gravity  darkening - and assumes  an edge-on  inclination, it
does suggest  that the observed bumps  and dumps in the  line profiles
are  not modulated by rotation, which  also  argues  against  the
existence  of spots  on the  surface  of the  star.  Similar  temporal
behaviour  of bumps  and dumps  crossing the  profiles of  photosperic
lines from blue  to red on time  scales of hours are  well known since
several     decades     in      other     $\delta$     Scuti     stars
\citep{walker87,kennelly92,kennelly98,balona00,mantegazza00},  as well
as in Be stars \citep[e.g.][]{vogt83,baade84,kambe93a,kambe93b,leister94},  
and early-type supergiants \citep[e.g.][]{baadeferlet84}. The  common
interpretation is that non-radial (longitudinal) pulsations are caused
by  motions along  longitudinal  strips;  those non-radial  pulsations
divide the stellar surface in  regions with different velocity fields,
which, in  the presence  of rotation, redistribute  the flux  over the
absorption line profile to create moving patterns of peaks and troughs
\citep[e.g.][and references therein]{walker87,telting97a,schrijvers99,mantegazza00,telting03}.  The
non-radial pulsations can  even be related to stellar  rotation in the
sense that  rapid rotation facilitates  the detection of this  kind of
pulsations   \citep{baadeferlet84}.   A   deeper   analysis   of   the
spectroscopic photospheric variability in  conjunction with the {\em TESS}
photometric data is deferred to another work.

\subsection{The circumstellar disk}

The common feature shared by  the detected CS absorptions mentioned in
Section  \ref{sec:variabilityofcsabsorptions}  is   the  $\sim$3  km/s
emission at  the core of the  lines.  We have checked  the spectra of
all   shell   stars  - 18 objects -   included   in   the   sample   of
\cite{rebollido20}    and    no     comparable  spectral characteristic is 
present, with the exception of a similar  stable Ca  {\sc ii}  
triplet profile with an emission at  the center and the blue side  weaker than 
the red side  in the stars HD 21688, HD 39182, and HD 42111. In the case of HD
21688 an  emission is  also observed  in the  center of  the H$\alpha$
line, while HD 39182 and HD 42111 display complex Fe {\sc ii} 5169 \AA
~profiles. At the same time, the shape of the Ca {\sc ii} triplet in 
HD 85905 is similar to the shape in these stars, but its radial velocity changes
remarkably as it happens in other shell features, while the 
photospheric lines remain constant. Further, we  note that all the  
previous mentioned stars but HD 21688
have variable  Ca {\sc  ii} H\&K  CS features, as  well as  that other
shell  stars  with   variability  in  these  Ca  {\sc   ii}  lines  in
\cite{rebollido20} - HD 256 (HR 10), HD 37306, HD 50241, HD 138629, HD 148283,
and HD 217782 - do have single narrow shell absorptions.
Thus, the emerging  picture of the A-type stars with  CS shells is the
one  of  a  heterogeneous  class   of  stars,  which  merely  reflects
previously known results \cite[e.g.][]{jaschek88,jaschek98}.

Interestingly,  alike the CS  shell  line  profiles in  $\phi$  Leo,
central emissions  at the  core of shell lines with  two absorption
minima   at   both   sides   are  also observed  in  the rapidly 
rotating Be   shell   stars \citep[e.g.][]{rivinius99}.  Such  
emissions, so-called  "central quasi emission" peaks, CQE's,  are
non-photospheric since  they are  only observed  in   shell  lines
\citep{koubsky93,koubsky97},   and  are favoured by the presence 
of an edge-on CS  disk, which should be optically thin in the continuum,  
have a small spatial  extent, and show  little line  broadening
\citep{hanuschik95,rivinius99}.  We note here that $\phi$ Leo also 
is a rapidly rotating star with a - plausible -  edge-on CS
disk. CQE's result  from  the local  minimum  at  zero  radial 
velocity  in  the fraction of the stellar disk occulted by CS gas 
in Keplerian orbital motion,  i.e., gas  moving perpendicular  
to the  line of  sight, as predicted  in Hanuschik's  Keplerian 
model  for CS  disks around  Be shell stars; thus,  CQE's are pure 
absorption  phenomena occuring at few   stellar  radii not related 
to   any   emission   process \citep{hanuschik95,rivinius99}. 
Following the model by \cite{hanuschik95} we can estimate the 
radius  of the $\phi$ Leo CS disk ($R_{\rm abs}$)  where the shell absorption 
lines form; thus, using Equation 1 in \cite{rivinius99} -
$R_{\rm abs} = (\Delta v_{\rm  cusps}/2v_{\rm crit} \sin i)^{2/3} $,
where $\Delta v_{\rm  cusps}$ is the velocity separation  of the two
minima around the  CQE, and $v_{\rm crit}$ is  the critical velocity
of the star ($v_{\rm crit} =  (G M_\star/R_\star)^{0.5}\simeq370$ km/s,
assuming the values of $ M_\star, R_\star$ given in Sect. 2 for 
$\phi$ Leo and considering  only the  centrifugal  and gravitational  
forces) -  we obtain $R_{\rm abs}$  equal to $\sim$13.5 $R_\star$ for  
the Ca {\sc ii}  K  line,  $\sim$10.7 $R_\star$ for  Ti {\sc ii} 3759  \AA,
$\sim$9.8  $R_\star$  for Fe  {\sc  ii}  5169  \AA, $\sim$8.5 $R_\star$
for Ca {\sc ii} 8542 \AA,  and $\sim$5.7 $R_\star$ for H$\alpha$. We
note  that  these values  of  $R_{\rm  abs}$  are similar  to  those
estimated  by \cite{abt97}  for the  Ti  {\sc ii}  lines assuming  a
Keplerian  scenario.  Variability in  the  strength  of the  absorption
minima at both sides of the CQE's, similar to the changes observed in
$\phi$ Leo, are also  known to be exhibited by some  Be shell stars -
e.g.  $\epsilon$  Cap  \citep[see  Fig.  3  in][]{rivinius99};  such
changes might be due to the re-supply of matter from the star to the
disk. Mechanisms for sufficient angular momentum   transfer   
from the star to the disk in Be stars are a matter of debate \citep[e.g. see][]{rivinius13}; 
among them, the more firmly established are those relating the disk and stellar non-radial pulsations \citep[e.g.][and references therein]{cranmer09}. 
Given the plausible evidence for the presence of this kind of pulsations in $\phi$ Leo, 
provided by the spectroscopic variations of its photospheric lines, 
it is suggestive to suppose that such mechanism could also be at work in
this star.

We also note that the absorption minima at both sides of the central emission 
always show up at similar radial velocities for each line, but different for 
each one, without any apparent dynamical evolution; thus differing from the 
bona-fide  exocometary events observed in $\beta$ Pic. This fact, together with 
the variability and changes observed from practically  one  spectrum  to 
another, at any time scale, in the lines of Ca {\sc ii}, Fe {\sc ii}, and 
Ti {\sc ii}, suggests that such line variations in $\phi$ Leo are highly 
unlikely to be produced by comet-like bodies.

\subsection{ A brief note on the exocomet-host shell stars}

As indicated in the Introduction, around 30 stars have been suggested to host
exocomets in their immediate surroundings \citep[][and references
therein]{welsh18,rebollido20,strom20}; this fact mainly based on the variable 
events observed in the Ca {\sc ii} K line. Among those 30 stars, 10 are A-type shell stars,
including $\phi$ Leo and HR 10 - see Table 7 in \cite{rebollido20}. However, the HR 10 results
presented by \cite{montesinos19} and those of $\phi$ Leo presented here, in both cases 
based on a very large amount of observational data, clearly demonstrate the 
need, at least in the shell stars, of a deeper analysis of the stellar spectra,
extending to large periods of time as well as to the behaviour of other shell 
lines, and not just restricting such analysis to the Ca {\sc ii} K line. For
instance, we note that \cite{rebollido20} pointed out in the case of the shell 
star HD 85905 that the observed spectral variations could be due to variability 
of the CS shell as suggested by the observed changes of several shell lines, 
and not just the Ca {\sc ii} K line; further, we have noted in the first 
paragraph of the previous section the similar or dissimilar behaviour of 
different lines among shell stars independently of the variability of the 
Ca {\sc ii} K line.

\section{Conclusions}
  
Our spectroscopic observations reveal that $\phi$ Leo is a highly variable
$\delta$ Scuti type  A-shell star with remarkable changes in its photospheric
and CS lines. The photospheric variability manifests as dumps and bumps 
superimposed on the lines profiles, varying their strength and sharpness and
propagating from blue- to more red-shifted  radial velocities; those features
persist along few hours with a time scale of the order of the $\delta$ Scuti
photometric variations observed with {\em TESS} \citep{balona20}. Thus, 
similarly to other $\delta$ Scuti stars, the  $\phi$ Leo spectroscopic
photospheric variability is likely produced by non-radial  pulsations caused by
motions along longitudinal strips on the stellar surface.

With respect to the non-photospheric CS lines, our data suggest that $\phi$ Leo 
is a late A-type star, whose shell is a disk seen nearly edge-on,  and with characteristics reminiscent of the ones observed in Be shell stars. Thus, we
would be facing a case of a {\em relatively late spectral type star, and
consequently relatively low mass star},  exhibiting a Be shell-like phenomenon.
Summarizing, $\phi$ Leo plausibly is a rapidly rotating $\delta$ Scuti  star
surrounded by a variable CS disk, possibly re-supplied by the $\delta$ Scuti
pulsations. To our knowledge, theoretical efforts focusing on such scenario have
been limited to B stars but not to late-type A stars \citep[e.g.][and references
therein]{rivinius13}; in this respect, $\phi$ Leo is an incentive to extend such
studies to less massive stars. Finally,   we  find  that this CS scenario for
$\phi$ Leo  is more plausible than the exocomet one first suggested by
\cite{eiroa16};  in fact, the central emission together with  
the variable absorption minima at both sides changing their strength, but not
their velocity,  in the lines of Ca {\sc ii}, Fe {\sc ii}, and Ti {\sc ii} from
practically  one spectrum  to another, and in some periods also in H$\alpha$,
are highly unlikely to be produced by comets.

Finally, the observational results presented in this work, together with those 
presented by \cite{montesinos19} concerning the shell star HR 10, suggest the need
of a critical revision regarding the origin of the Ca {\sc ii} K line variable
features observed in other shell stars, which have been attributed to exocomets.  

\begin{acknowledgements}
The authors thank the referee for the comments and suggestions to the original  manuscript. This work is based on observations made at the Spanish Observatorio del Roque de los Muchachos (La Palma, Spain) of the Instituto de Astrofísica de Canarias with the Mercator Telescope, operated by the Flemmish Community, the Nordic Optical Telescope, operated by the Nordic Optical Telescope Scientific Association, and the Italian Telescopio Nazionale Galileo (TNG) operated  by the Fundación Galileo Galilei of the INAF. Based on observations made with the ESO/MPG 2.2 m telescope at the La Silla Observatory under programmes 099.A-9029(A) and 094.A-9012. CE, BM, IR, GM and EV  were supported by Spanish grant AYA 2014-55840-P; BM, IR, GM and EV are also supported by Spanish grant PGC2018-101950-B-100. We thank Ana Guijarro of the Calar Alto Observatory for obtaining the CARMENES spectrum. 
\end{acknowledgements}

\bibliographystyle{aa} 
\bibliography{hr4368}

\begin{appendix}

 \section{Results: Photospheric lines}
\label{Appendix A}

\subsection{HERMES May 2016}

\begin{figure}[!ht] 
  \centering
\scalebox{0.3}{\includegraphics[angle=-90]{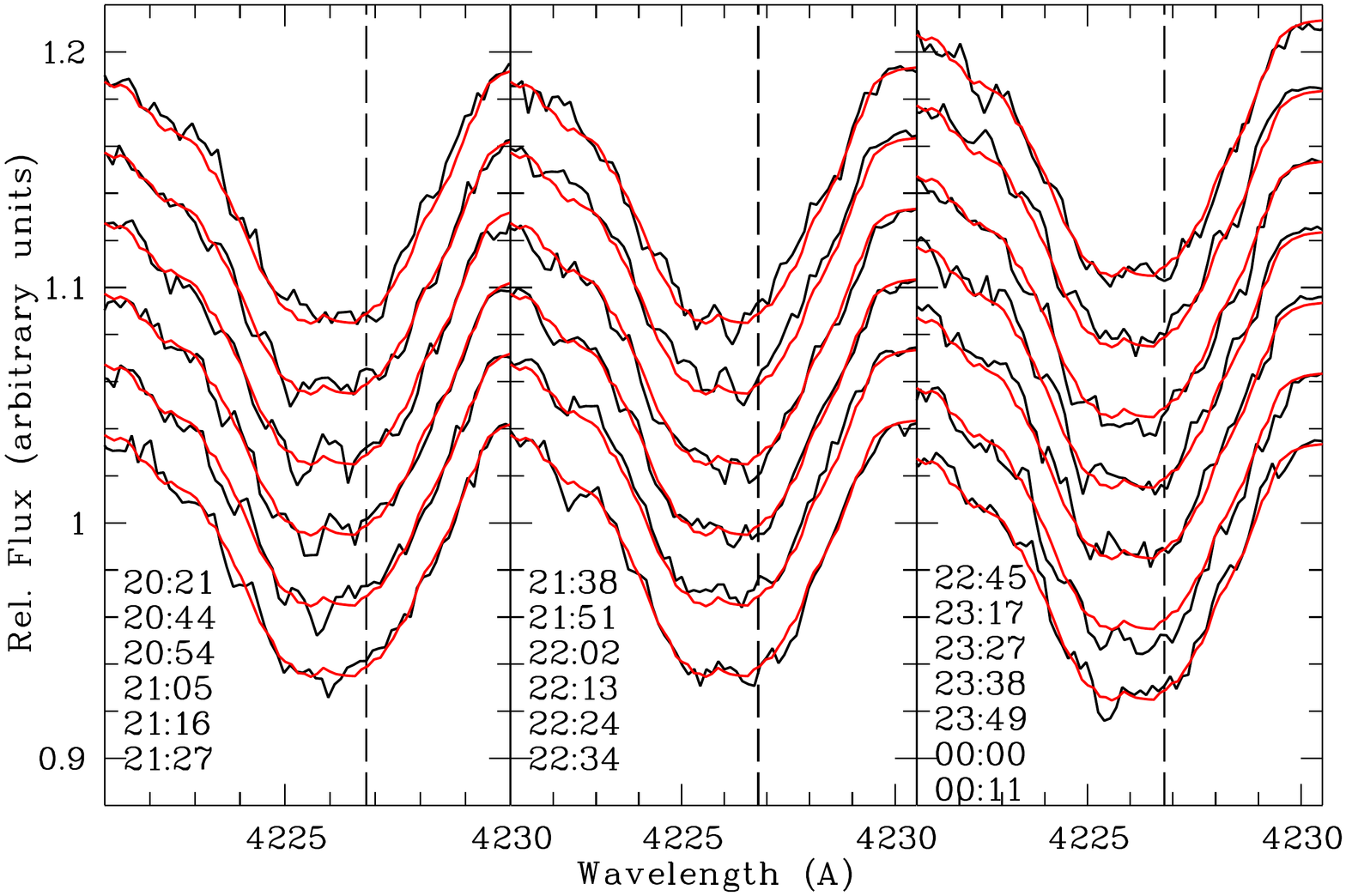}}
\scalebox{0.3}{\includegraphics[angle=-90]{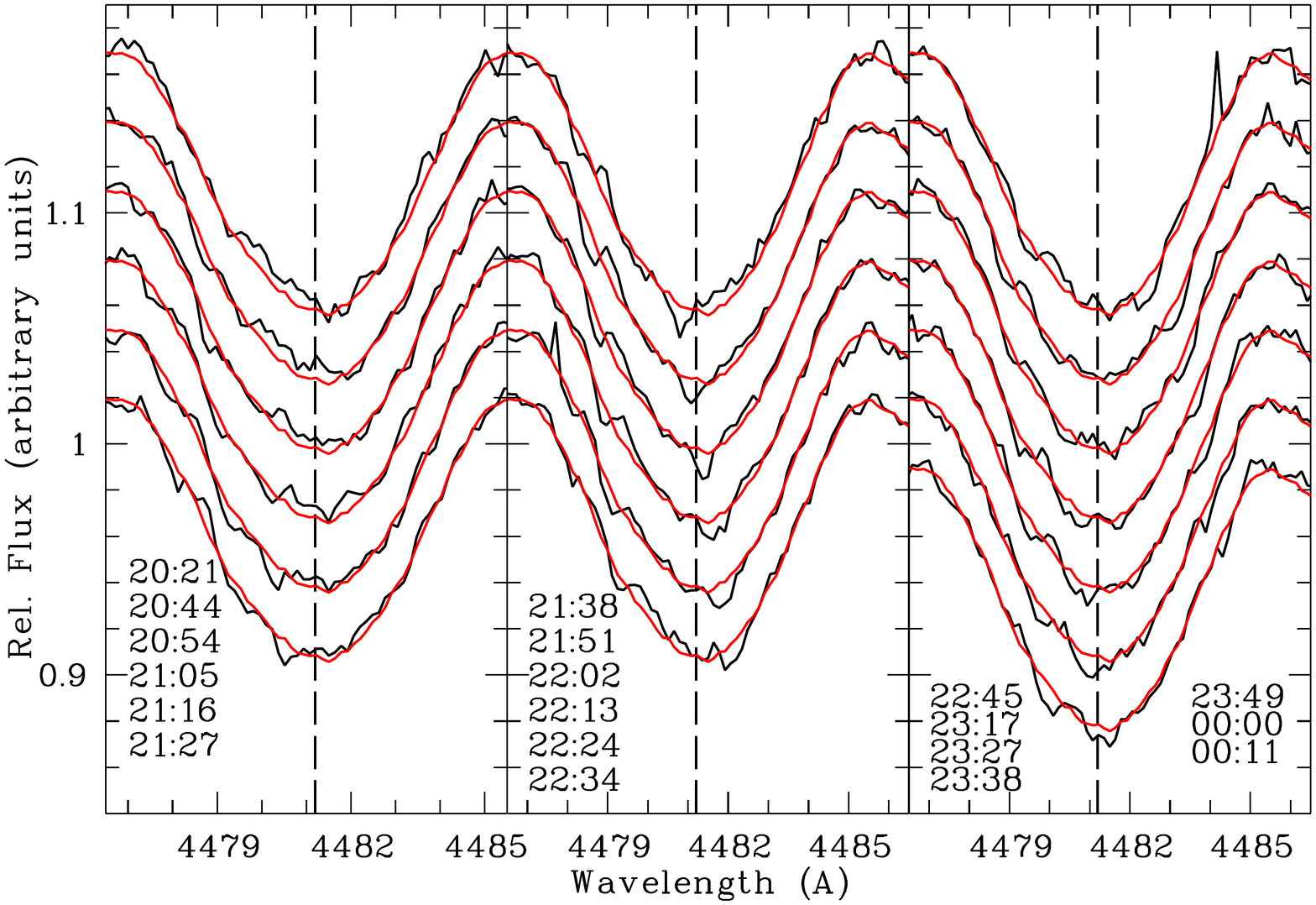}}
\caption{Profiles of the Ca {\sc i} 4226 \AA ~(top) and  Mg {\sc ii} 4481 \AA  ~(bottom) lines as observed with HERMES in 2016-05-11. UTs from top to bottom and from left to right are indicated in the labels.
Spectra have been rebinned to 0.1406 \AA/pixel.  The median of all spectra superimposed on each spectrum is plotted in red. }
\label{201605_cai_mgii} 
\end{figure}

During the night 2016-05-11, 20 consecutive spectra were obtained along $\sim$4 hours of observations. The night was not optimal and one of the spectra was not useful. Fig. \ref{201605_cai_mgii} shows the profiles of Ca {\sc i} 4226 \AA ~and the Mg {\sc ii} lines obtained during this night. Several dumps/bumps lasting up to several hours are observed in both lines but seemingly not coincident. In particular, e.g., there is a depression in the Mg {\sc ii} line going from $\sim$4480 \AA ~to 4483 \AA ~from $\sim$ 2016-05-11 21h05m UT to at least 2016-05-11 23h27m UT, i.e., it lasts at least $\sim$ 150 minutes

\subsection{HERMES/FEROS March/April 2017}

During these observing runs series spectra were taken  during  12 consecutive nights as long as the air mass of the star was $\lesssim$2.0. In some nights HERMES and FEROS spectra were obtained simultaneously and their appearance was quite similar in spite of the different spectral resolution, which provides high confidence about the reliability on the weak features seen superimposed on the line profiles. As representative results we show below the spectra of the Mg {\sc ii} 4481 \AA ~line in two of the nights, in one case with HERMES spectra, and one night with both HERMES and  FEROS spectra.

\begin{itemize}
\item {\bf 2017-03-28}: During this night 24 spectra were obtained with HERMES (2 of them were not useful).  Fig \ref{20170328_mgii} plots the Mg {\sc ii} 4481 \AA ~line with the labels indicating the observing UT of each spectra. No obvious dumps/bumps in the line profiles that could be attributed to pulsations are clearly detected in several spectra, although very weak  features might be present in some of them.

\begin{figure}[!ht] 
  \centering
\scalebox{0.3}{\includegraphics[angle=-90]{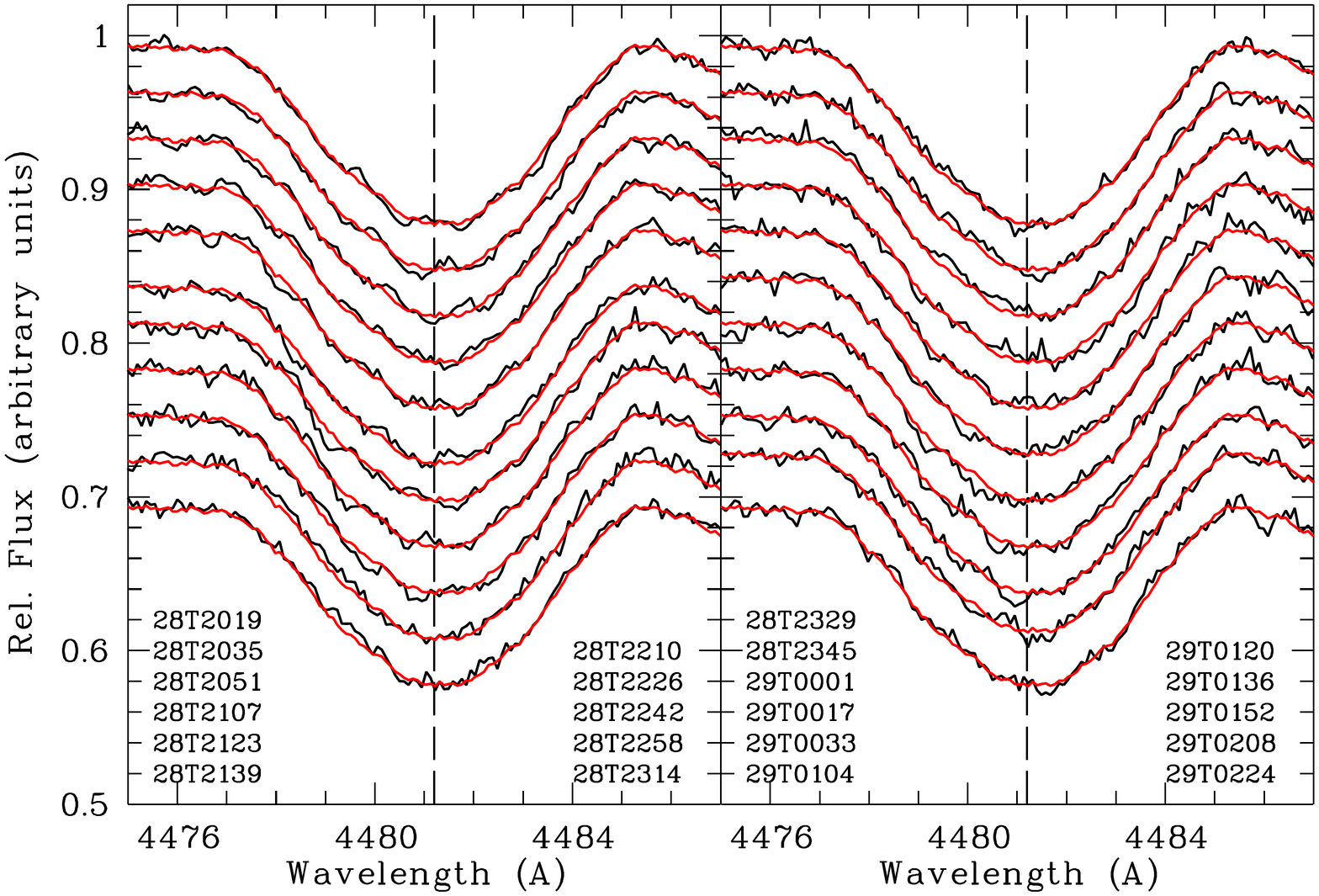}}
\caption{Profiles of the Mg {\sc ii} 4481 \AA ~line as observed with
HERMES during the night 2017-03-28/29. The spectra have been rebinned
to 0.09 \AA/pixel; the median of the night is plotted in red 
superimposed on each spectrum. In both panels UTs from top to bottom and from left to right correspond to spectra ordered from top to bottom.}
\label{20170328_mgii} 
\end{figure}  

\item {\bf 2017-04-01/02}:  This night 21 HERMES and 28 FEROS spectra were taken along a time interval of $\sim$9:30 hours. Fig. \ref{20170401_mgii} plots the Mg {\sc ii} 4481 \AA ~line of these spectra. Several features, mostly dumps, are distinctly present in the spectra, although  no obvious features are discernible in some of them (e.g. the spectra taken at  2017-04-02 04h23m UT or 2017-04-02
04h28m UT). Note that some features appear in the blue side of the line profile but not always apparently propagate to the red wing. It is noticeable that the HERMES and FEROS spectra agree very well, see e.g the spectra taken at  2017-04-02 01h44m UT and 2017-04-02 01h45m UT with HERMES and FEROS respectively, or the ones taken  at 2017-04-02 02h32m UT with both HERMES and FEROS simultaneously.
\begin{figure*}[ht] 
  \centering
\scalebox{0.3}{\includegraphics[angle=-90]{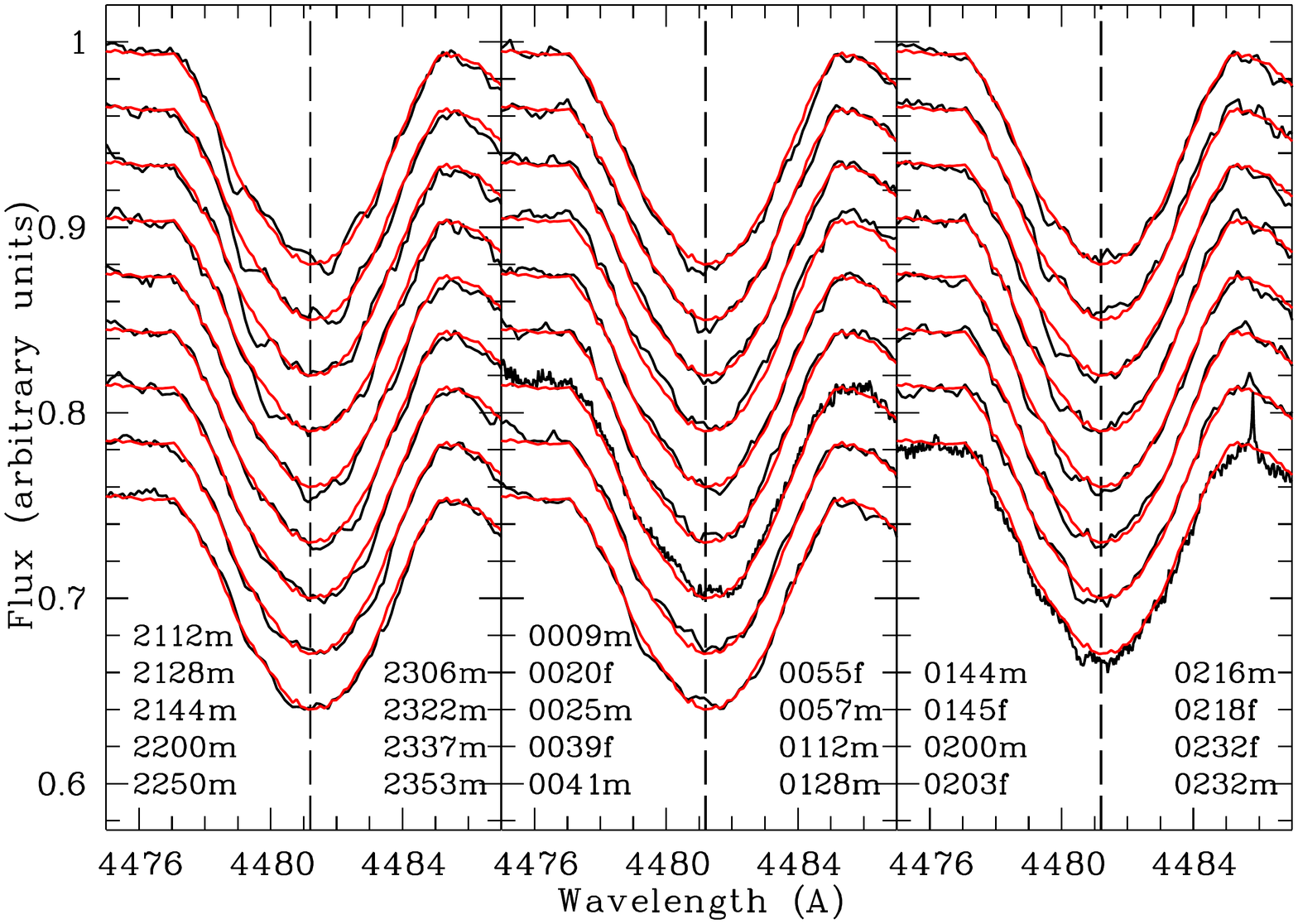}}
\scalebox{0.3}{\includegraphics[angle=-90]{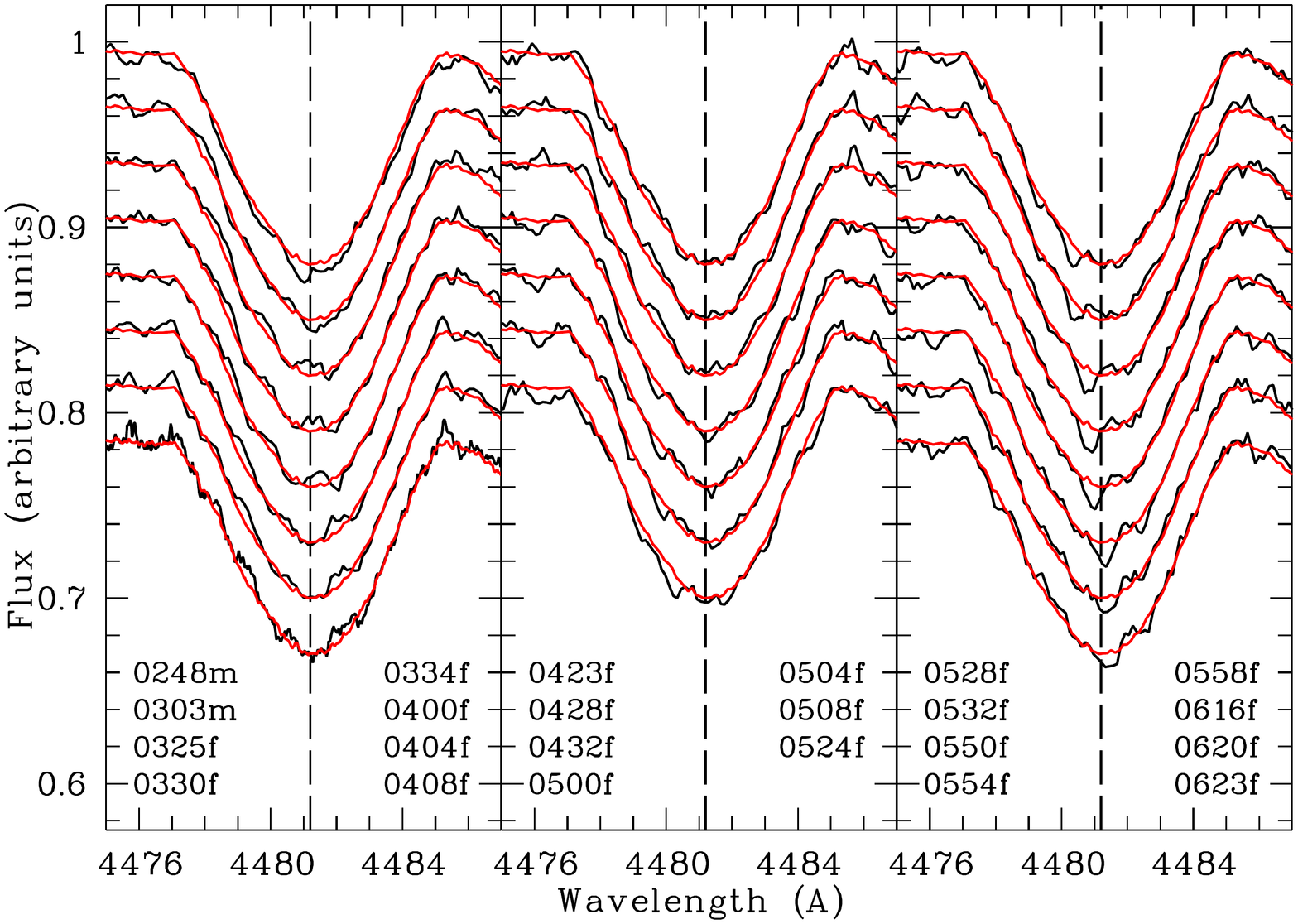}}
\caption{Profiles of the Mg {\sc ii} 4481 \AA ~line as observed with HERMES and FEROS during the night of 2017-04-01/02 along $\sim$9:30 hours. All spectra have been rebinned to 0.12 \AA/pixel, and the median of the night is plotted in red  superimposed on each spectrum. Spectra with ``m'' in the labels refer to those obtained with HERMES and those with ``f'' to FEROS.
Labels correspond to spectra  from top to bottom and then the two columns from left to right.}
\label{20170401_mgii} 
\end{figure*}

\end{itemize}

\section{Results: Non-photospheric lines}

\subsection{HERMES March 2017}

\begin{figure*}[!ht]
  \centering
\scalebox{0.25}{\includegraphics[angle=-90]{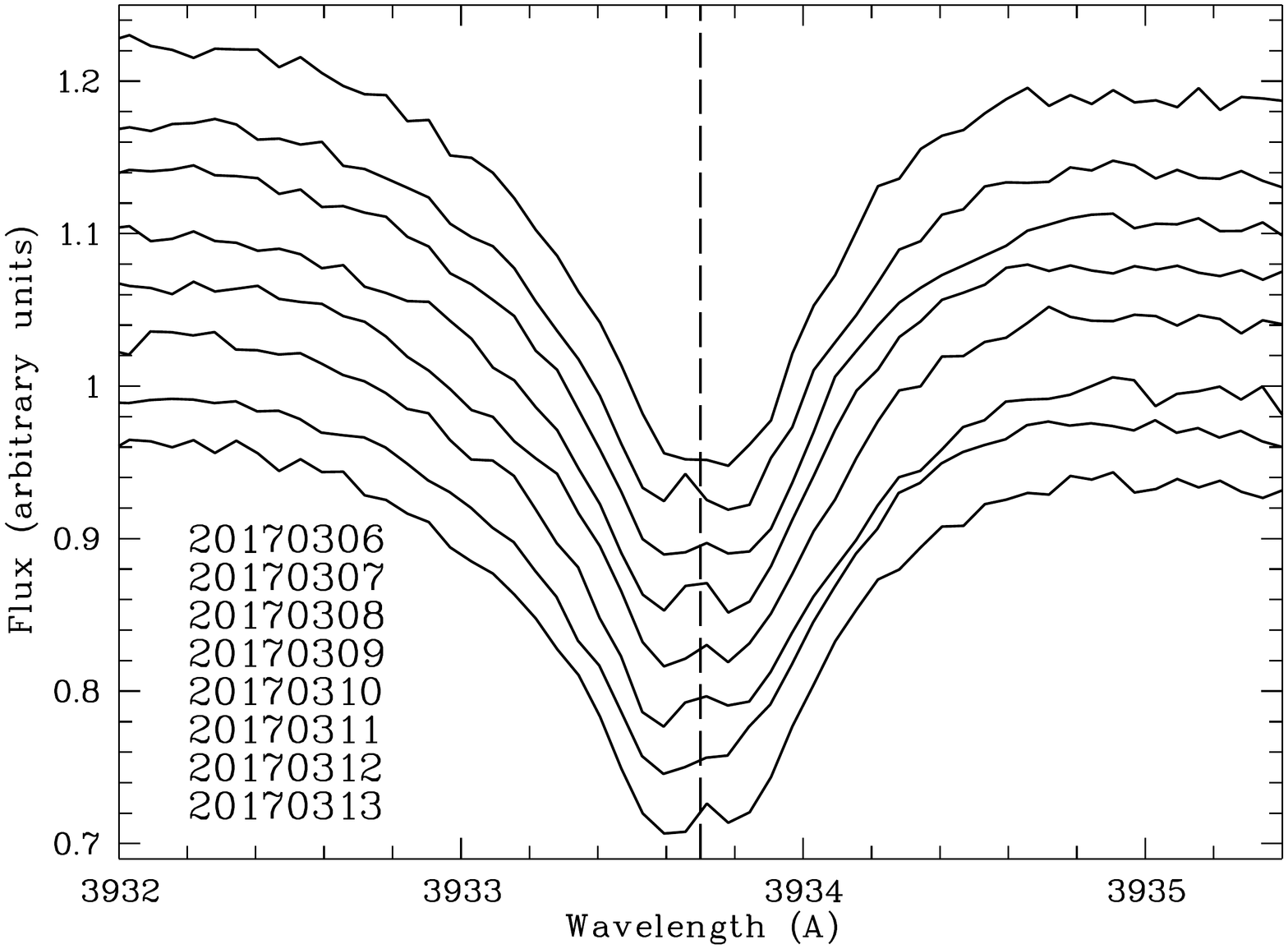}}
\scalebox{0.25}{\includegraphics[angle=-90]{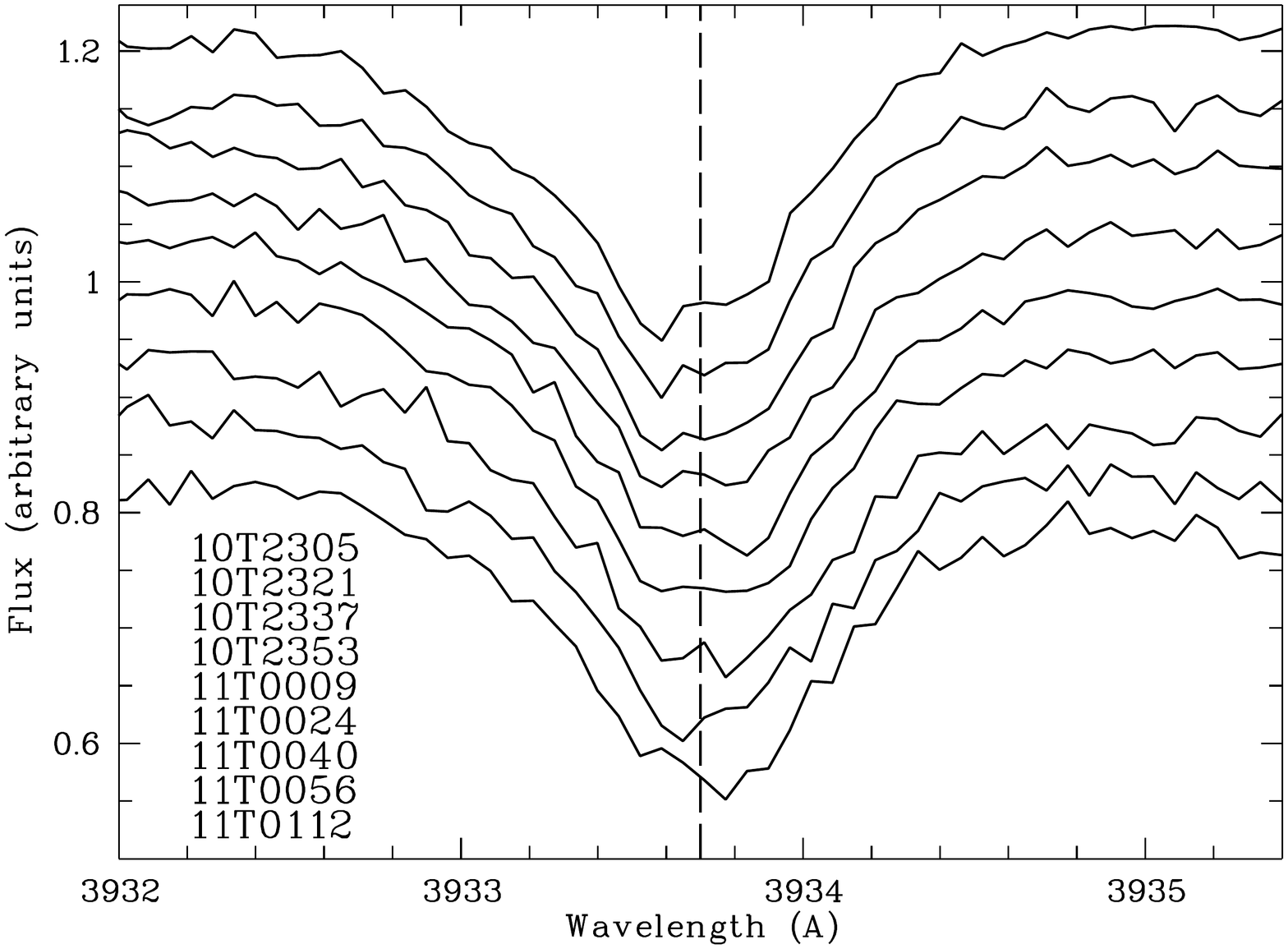}}
\scalebox{0.25}{\includegraphics[angle=-90]{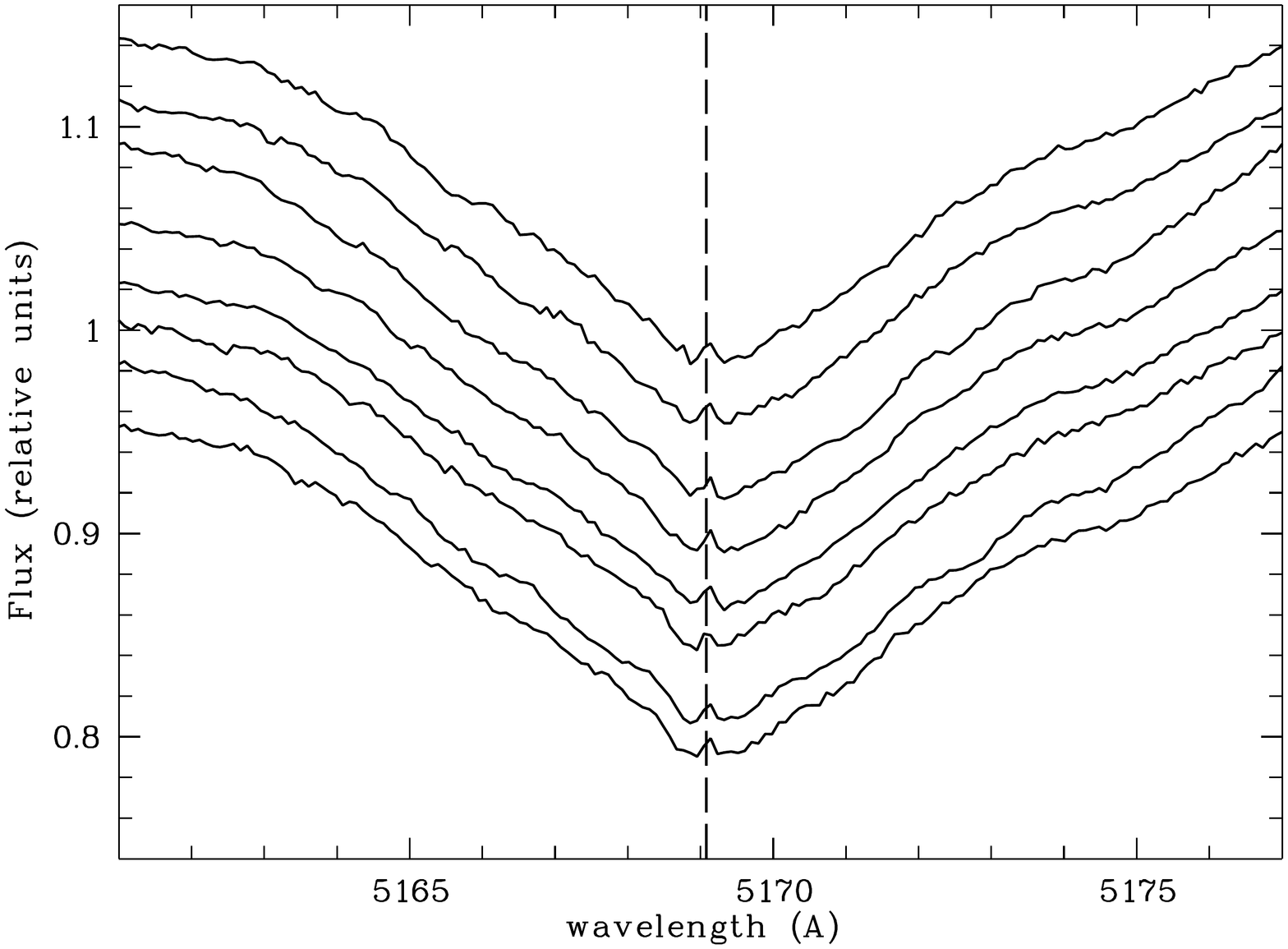}}
\scalebox{0.25}{\includegraphics[angle=-90]{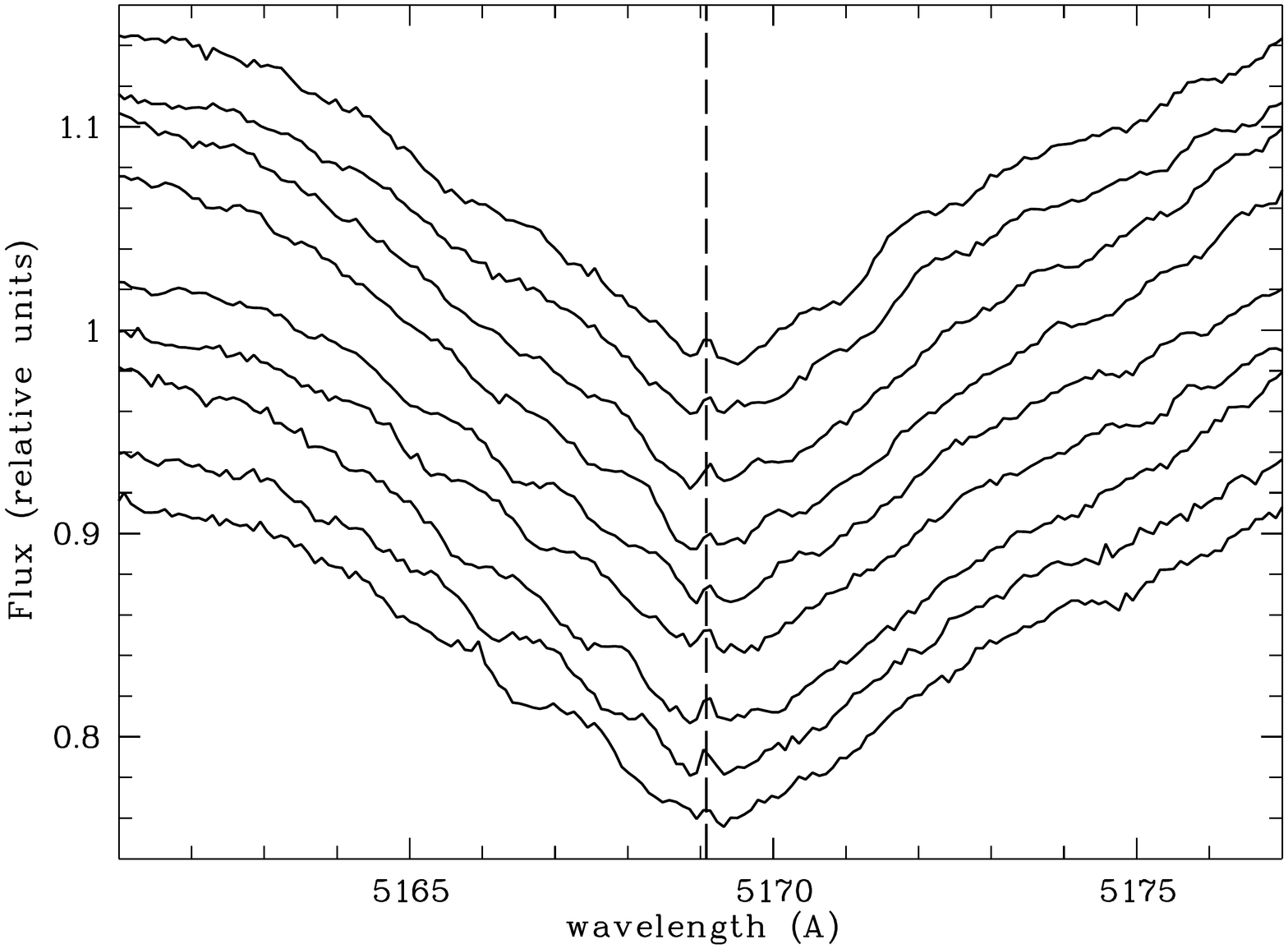}}
\caption{Left:  Nightly median  of Ca  {\sc ii}  K  and  Fe  {\sc ii}  5169 \AA.
  Dates are indicated  in the labels. Right: Some consecutive spectra  taken during  the 
  night  10/11 March  2017. Exposure time of each spectrum is 15 minutes. The 
  resolution  of the Ca {\sc ii} spectra is is   0.0625  \AA/pixel, while for  
  Fe {\sc ii}  is  0.0938 \AA/pixel.  Ca  {\sc ii} K  shows variability
 from night to  night as well as irregular  short timescale variability
in individual  nights.  The central  emission is  seen in some  of the
nights, and both blue- and red-shifted minima change their relative
strength. With respect to the Fe {\sc ii} line the central emission is
present in all nights but the  variability is less pronounced than the
one in the K line and no correlation is seen.}
\label{mercator_201703} 
\end{figure*}

During this period with HERMES spectra were taken along 8 consecutive nights. 
Fig. \ref{mercator_201703} shows the median spectra corresponding to each night 
of Ca {\sc ii} K and Fe {\sc ii} 5169 \AA, as well as the spectra 
obtained during the night 2017-03-10. Variability is observed in a daily basis
and in consecutive spectra of a single night. Note that the changes of Ca 
{\sc ii} and Fe {\sc ii} are different, and that the 3 km/s emission is not
distinguishable in some spectra. 

\subsection{FEROS 2017 April}
\begin{figure*}[!ht]
  \centering
  \scalebox{0.2}{\includegraphics[angle=-90]{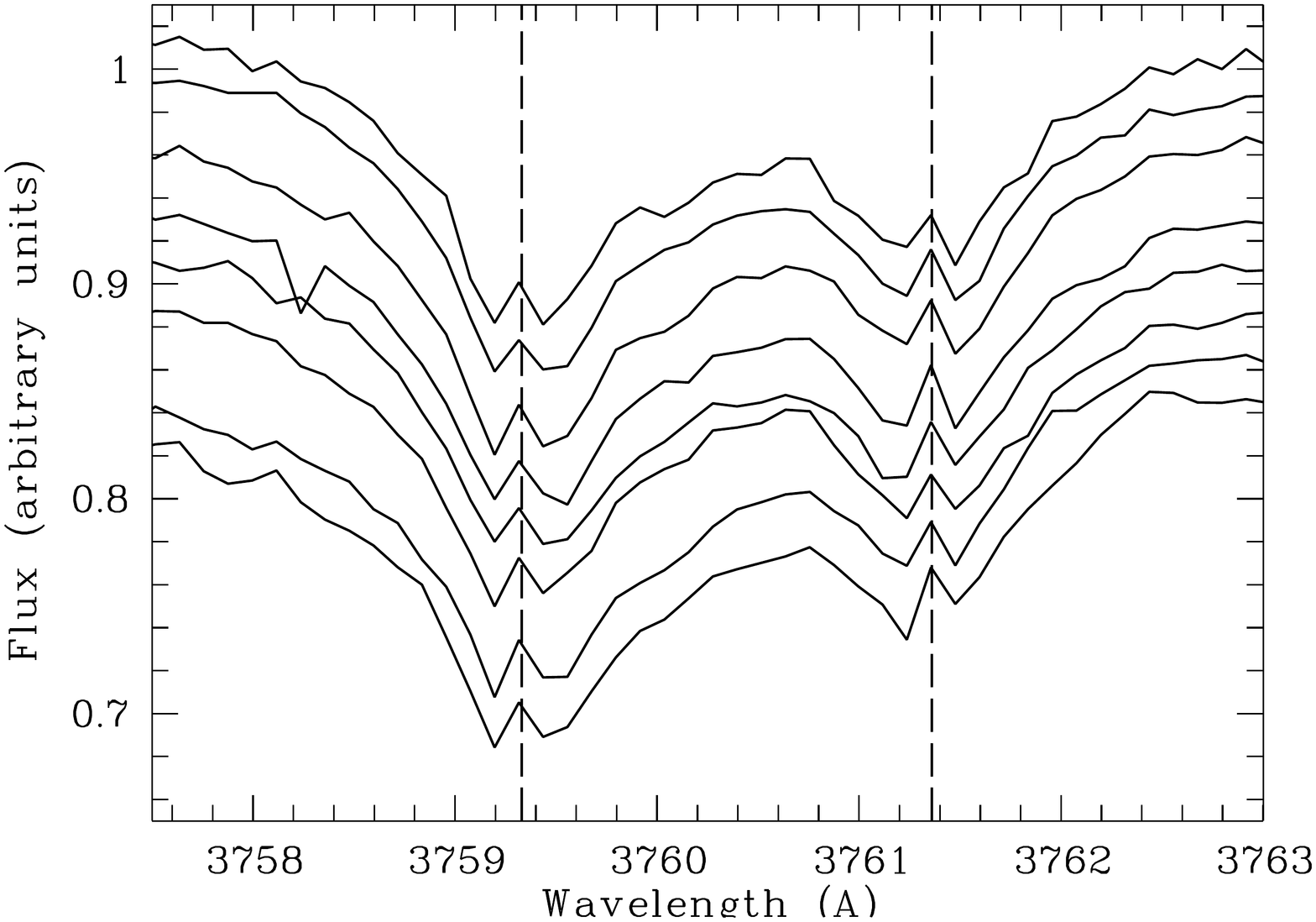}}
  \scalebox{0.2}{\includegraphics[angle=-90]{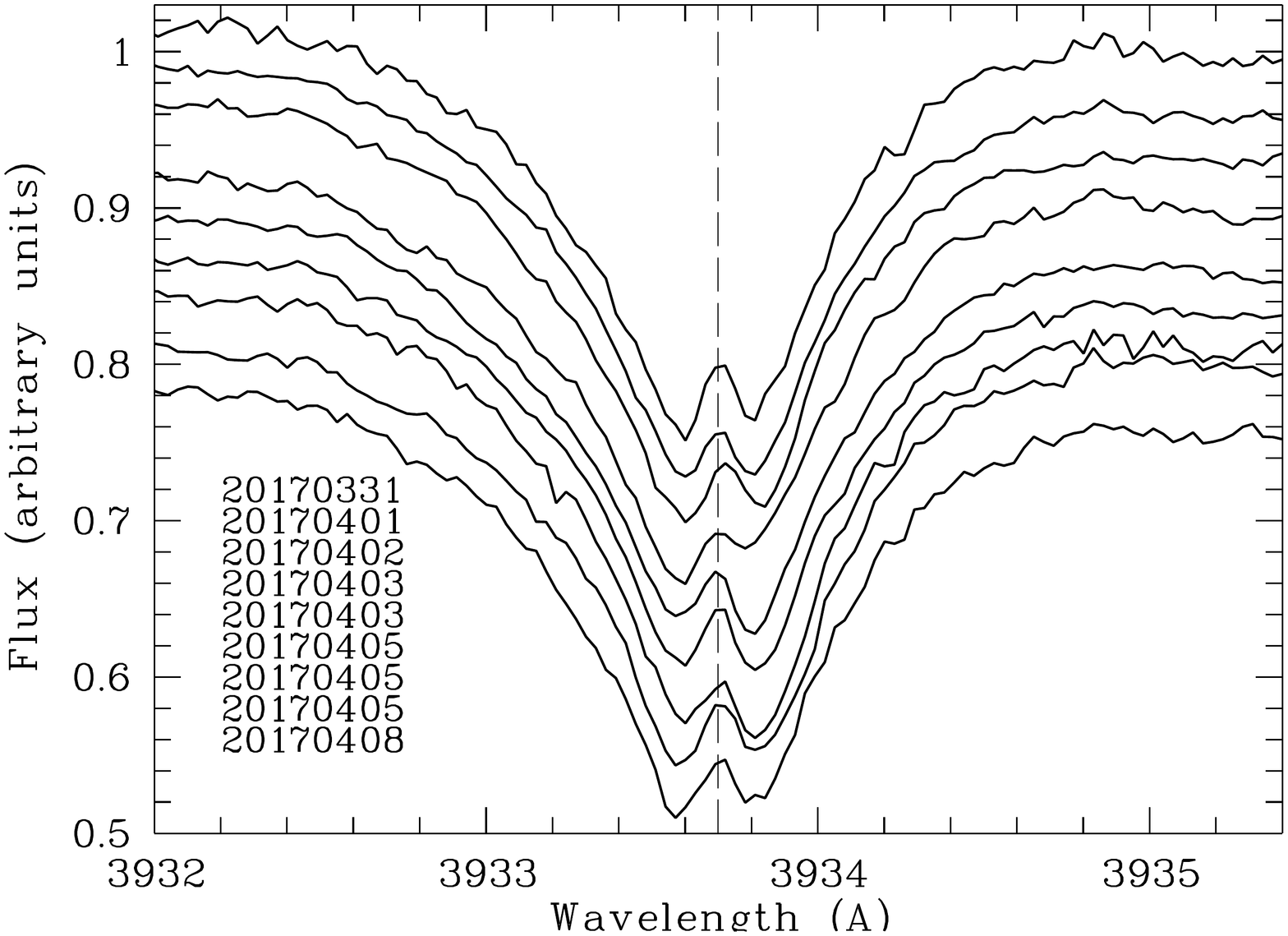}}
   \scalebox{0.2}{\includegraphics[angle=-90]{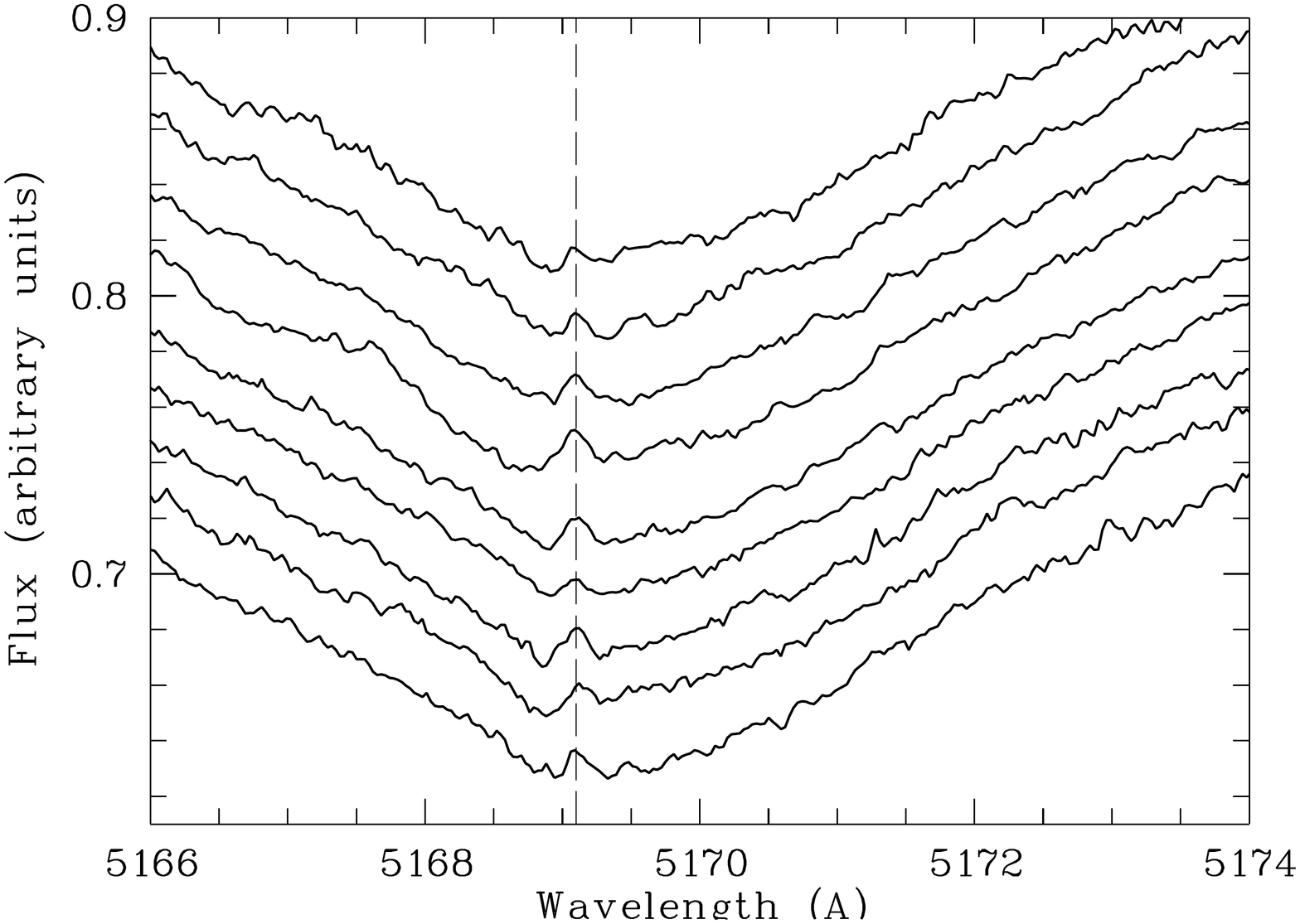}} 
\caption{Nightly median of Ti {\sc ii} 3759 and 3761 \AA, Ca {\sc ii} K and Fe {\sc ii} 5169 \AA ~lines as observed from 20170331 to  20170408 with FEROS. Dates are indicated from top to bottom in the Ca  {\sc ii} K panel.}
\label{feros_201704_medians} 
\end{figure*}

During this period spectra were taken along 9 consecutive nights with 
FEROS. Fig. \ref{feros_201704_medians} shows the median spectra of all nigths
corresponding to Ti {\sc ii} 3759 and 3761 \AA, Ca {\sc ii} K, and Fe {\sc ii} 
5169 \AA. We note that the 3 km/s emission is observed in all nights and 
that the variations are different from line to line.

\section{Observing Log}
\label{Appendix_Observing_Log}
Table C.1 is published is only available in electronic form
at the CDS via anonymous ftp to cdsarc.u-strasbg.fr (130.79.128.5)
or via http://cdsweb.u-strasbg.fr/cgi-bin/qcat?J/A+A/

\end{appendix}
\end{document}